# Thulium anomalies and rare earth element patterns in meteorites and Earth: Nebular fractionation and the nugget effect

Nicolas Dauphas [a] and Ali Pourmand [a, b]

[a] Origins Laboratory, Department of the Geophysical Sciences and Enrico Fermi Institute, The University of Chicago, 5734 South Ellis Avenue, Chicago IL 60637, USA

[b] Neptune Isotope Laboratory, Department of Marine Geosciences, The University of Miami – RSMAS, 4600 Rickenbacker Causeway, Miami, FL 33149, USA

[*] Corresponding author e-mail address: dauphas@uchicago.edu

Tel: 1-773-702-2930

**Abstract:** This study reports the bulk rare earth element (REEs, La-Lu) compositions of 41 chondrites, including 32 falls and 9 finds from carbonaceous (CI, CM, CO and CV), enstatite (EH and EL) and ordinary (H, L and LL) groups, as well as 2 enstatite achondrites (aubrite). The measurements were done in dynamic mode using multi-collector inductively coupled plasma mass spectrometers (MC-ICPMS), allowing precise quantification of mono-isotopic REEs (Pr, Tb, Ho and Tm). The CI-chondrite-normalized REE patterns ($La_N/Lu_N$; a proxy for fractionation of light *vs.* heavy REEs) and Eu anomalies in ordinary and enstatite chondrites show more scatter in more metamorphosed (petrologic types 4 to 6) than in unequilibrated (types 1-3) chondrites. This is due to parent-body redistribution of the REEs in various carrier phases during metamorphism. A model is presented that predicts the dispersion of elemental and isotopic ratios due to the nugget effect when the analyzed sample mass is limited and elements are concentrated in minor grains. The dispersion in REE patterns of equilibrated ordinary chondrites is reproduced well by this model, considering that REEs are concentrated in 200 μm-size phosphates, which have high $La_N/Lu_N$ ratios and negative Eu anomalies.

Terrestrial rocks and samples from ordinary and enstatite chondrites display negative Tm anomalies of ~-4.5 % relative to CI chondrites. In contrast, CM, CO and CV (except Allende) show no significant Tm anomalies. Allende CV chondrite shows large excess Tm (~+10 %). These anomalies are similar to those found in group II refractory inclusions in meteorites but of much smaller magnitude. The presence of Tm anomalies in meteorites and terrestrial rocks suggests that either (*i*) the material in the inner part of the solar system was formed from a gas reservoir that had been depleted in refractory dust and carried positive Tm anomalies or (*ii*) CI chondrites are enriched in refractory dust and are not representative of solar composition for refractory elements. A new reference composition relevant to inner solar system bodies (CI*) is calculated by subtracting 0.15 % of group II refractory inclusions to CI. The observed Tm anomalies in ordinary and enstatite chondrites and terrestrial rocks, relative to carbonaceous chondrites, indicate that material akin to carbonaceous chondrites must have represented a small fraction of the constituents of the Earth. Tm anomalies may be correlated with Ca isotopic fractionation in bulk planetary materials as they are both controlled by addition or removal of refractory material akin to fine-grained group II refractory inclusions.

# 1. Introduction

Lanthanides (La, Ce, Pr, Nd, unstable Pm, Sm, Eu, Gd, Tb, Dy, Ho, Er, Tm, Yb and Lu), also known as rare earth elements (hereafter REEs), have diagnostic signatures associated with a variety of geochemical and cosmochemical processes. Trivalent REEs fractionate during these processes as smooth functions of their masses, a phenomenon known as lanthanide contraction whereby the ionic radii of REEs decreases with increasing atomic masses. Exceptions are Ce, which can also exist as 4+ under oxidizing conditions, and Eu (and to some extent Yb), which can exist as 2+ under reducing conditions. These peculiar valence states can lead to deficits or enrichments in Ce and Eu relative to neighbor REEs in planetary materials. Meteorites and their constituents display variations in REE concentrations and abundance patterns that reflect evaporation/condensation processes in the nebula, parent-body aqueous alteration, metamorphism, magmatic differentiation, and terrestrial weathering during residence on the Earth's surface. A major difficulty is to disentangle these various processes in bulk meteorite compositions.

Calcium aluminum inclusions (CAIs) in primitive chondrites were formed by condensation of solar nebula gas and exemplify the kinds of geochemical variations that were produced during the earliest stages of solar system evolution. In particular, their REE patterns, normalized to the mean of CI-chondrites, show significant departures from the smooth trends discussed above that can only be explained if these elements were fractionated from one another during evaporation/condensation processes (Boynton, 1975). Martin and Mason (1974) classified REE patterns of CAIs into 6 groups, one of which (group II) had first been identified by Tanaka and Masuda (1973) and is characterized by depletion in the most refractory and volatile REEs. This pattern is common in CAIs, representing ~1/3 of the 283 CAIs surveyed by Fegley and Ireland (1991), and is thought to represent a snapshot in the condensation sequence of the REEs (Boynton, 1975; Davis and Grossman, 1979; Kornacki and Fegley, 1986).

During evaporation and condensation processes in a gas of solar composition, Eu and Yb are more volatile than LREEs (and Tm), which in turn are more volatile than ultrarefractory REEs Gd-Er and Lu. The most refractory REEs (Gd-Er and Lu) are depleted in group II REE pattern, presumably because they were removed as part of an ultra-refractory component that is missing from group II CAIs. The most volatile REEs (Eu and Yb) are also depleted in group II REE pattern because they never fully condensed. A telltale signature of this pattern is the presence of a positive Tm anomaly arising from the fact that Tm, a heavy REE (HREE), has similar volatility as light REEs (LREEs). The missing ultra-refractory component was originally proposed to be perovskite (Davis and Grossman, 1979) but it may as well be hibonite (MacPherson and Davis, 1994), which was identified in a few CAIs from Murchison and Allende that are highly enriched in ultra-refractory oxides (Boynton et al., 1980, Palme et al., 1982; Ireland et al., 1988, Simon et al., 1996, 2002; El Goresy et al., 2002 and Hiyagon et al., 2003). A group II REE pattern was identified in bulk samples of the Allende CV chondrite (Nakamura, 1974, Jarosewich et al., 1987, Shinotsuka et al., 1995, Shinotsuka and Ebihara, 1997, Makishima and Nakamura, 2006, Pourmand et al. 2012, Barrat et al., 2012; Stracke et al., 2012) but it is unknown whether this signature is present in other meteorites that contain a lower proportion of CAIs.

Prior to the advent of plasma mass spectrometry, the majority of REE data in meteorites were obtained by techniques such as Spark Source Mass Spectrometry (SSMS, Morrison et al., 1970, Martin and Mason, 1974 and Jarosewich et al., 1987) and Neutron Activation Analysis (NAA, Schmitt et al., 1963, Haskin et al., 1966, Morrison et al., 1970, Ebihara and Honda, 1984 and Jarosewich et al., 1987) that did not provide sufficient precision to detect Tm anomalies in bulk samples. Other techniques such as isotope dilution mass spectrometry (IDMS) provided high-precision data for multi-isotope elements but measurements could not be made on mono-isotopic Pr, Tb, Ho and Tm (Nakamura and Masuda, 1973, Nakamura, 1974, Evensen et al., 1978). We have developed a new technique to measure the abundance of actinides and REEs using multi-collector MC-ICPMS (Dauphas and Pourmand, 2011; Pourmand et al., 2010, 2012) that allows evaluating whether Tm anomalies produced in the nebula are present in bulk planetary materials. To that end, we analyzed the bulk REE compositions of 43 meteorite specimens (34 falls and 9 finds).

Variations in other REEs have also been documented in bulk chondrites, such as Eu anomalies or fractionation of LREE relative to HREE (Nakamura and Masuda 1973; Nakamura, 1974; Evensen et al., 1978) that could conceivably have a nebular origin. However, some of these variations show more dispersion in metamorphosed compared to unequilibrated chondrites, suggesting that they also arise from parent-body processes. REEs tend to be concentrated in minor phases (phosphate and oldhamite in carbonaceous and enstatite chondrites, respectively).

Given the limited mass of meteorite usually available for analysis, REEs are prone to the nugget effect, meaning that the concentration of phosphate and oldhamite could vary randomly from sample to sample. We present a model to predict the dispersion of elemental or isotopic ratios produced by the nugget effect and compare the theoretical predictions with the observations to assess whether those variations have a nebular or parent-body origin.

2. Materials and Methods

The analysis of REE abundances in bulk meteorites involved sample digestion by high-purity alkali flux fusion, TODGA extraction chromatography for matrix-analyte separation, and multi-collection plasma source mass spectrometry with desolvating nebulizers to minimize oxide interferences. A brief description is included in the following and additional details can be found in Pourmand and Dauphas (2010) and Pourmand et al. (2012).

ACS grade hydrochloric (HCl) and nitric ($HNO_3$) acids were distilled in sub-boiling quartz and PTFE Teflon stills. Concentrated acids were titrated with certified 0.1 and 1 mol $L^{-1}$ sodium hydroxide solutions and dilutions were prepared with high-purity water from a Millipore Milli-Q system (resistivity > 18 MΩ $cm^{-1}$). Teflon (PFA) vials were first cleaned in 50 % $HNO_3$ at 70 °C, followed by boiling in aqua regia (HCl:$HNO_3$ at 3:1). Platinum and quartz evaporation dishes were cleaned with hot, 4 mol $L^{-1}$ HCl and aqua regia, respectively. Pre-packed, 2-mL TODGA cartridges (resin grain size: 50-100 μm), connectors and a plexiglas vacuum chamber were purchased from Eichrom Inc. High-purity, 8-mL graphite crucibles, LiBr non-wetting agent (Pure grade) and certified multi-element REE standard solutions were obtained from SPEX CertiPrep. The SPEX CertiPrep multi-element Solution 1 (Lot# 40-54AS) contains all REEs plus Sc and Y. The certified concentrations of the standard for REEs are (in μg/g) 9.88 La, 10 Ca, 10 Pr, 9.96 Nd, 9.88 Sm, 10 Eu, 9.97 Gd, 10 Tb, 10 Dy, 9.99 Ho, 9.96 Er, 9.91 Tm, 10 Yb, 9.94 Lu, all calibrated against NIST SRM mono-elemental concentration standard solutions. The uncertainty on these concentrations is listed as ±0.5 %. Puratronic lithium metaborate ($LiBO_2$) was obtained from Alfa Aesar (99.997% metals basis, cat# 10739).

Allende CV3 reference material (Smithsonian Institution, NMNH 3529 split 8, position 5, Jarosewich et al., 1987) and Orgueil CI chondrite were received in powder and crumbs, respectively, and were processed through the chemistry without further treatment. Other meteorite pieces were examined for signs of surface alteration and fusion crust and if necessary, the affected areas were removed using a dry saw. To remove the residual powder from sawed surfaces, samples were cleaned with high-purity ethanol in an ultrasonic bath for approximately 10 seconds and dried under a heat lamp. The cleaned pieces were thoroughly homogenized using sand-cleaned agate mortar and pestle under class-100 laminar flow. Powder aliquots weighing 0.03-0.16 g were digested by flux fusion. Due to limited sample availability, the starting masses for the majority of meteorites ranged between 0.1-0.6 g but we also processed larger pieces (0.8-1 g) of Murchison (CM2), Indarch (EH4), Pillistfer (EL6), Ochansk (H4), Farmington (L5), Isoulane (L6), Kelly (LL4) and Paragould (LL5) to examine the effect of sample heterogeneity on REE abundance patterns and anomalies. Most samples were taken from meteorite fragments that had been analyzed previously by Dauphas and Pourmand (2011) for Lu-Hf-U-Th systematics.

High-temperature fusion with $LiBO_2$ flux was preferred over traditional hotplate or Parr Bomb dissolution techniques to ensure complete sample digestion. Alternative acid dissolution techniques take longer, involve multiple dry-down and acid conversion steps, and may result in incomplete dissolution of refractory minerals. We previously measured the concentrations of REEs in high-purity alkali flux from different vendors (Pourmand et al., 2012) and showed that commercially available flux of highest purity is still contaminated with light rare earth elements (LREE, La-Sm) and Yb. In order to process bulk meteorite samples with typically low REE concentrations, it is necessary to purify the commercial alkali flux further by passing it through an array of TODGA extraction chromatography resin cartridges. High-purity flux was recovered through sequential evaporation of the eluent in platinum and quartz crucibles and PFA transfer beakers (see Pourmand et al., 2012 for details).

The homogenized meteorite powder was fused with purified $LiBO_2$ at a sample to flux ratio of 1:6. Approximately 0.06-0.15 g of a non-wetting agent (LiBr) was also added and the mixture was fused in capped graphite crucibles at 1070 °C for 12 minutes. Addition of LiBr solution was critical to prevent the molten beads from adhering to the crucible and facilitate quantitative recovery of the fusion mix. The melt was subsequently poured into a 30 mL PFA Savillex beaker containing 15 mL of 3 mol L$^{-1}$ HNO$_3$. The shattered glass was dissolved within minutes on a Vortex at 6000 rpm after adding 10 mL of 3 mol $L^{-1}$ $HNO_3$ to the solution. Procedural blanks were processed in the same manner using purified flux only. While incomplete transfer of fusion melt can result in lower apparent concentrations of REEs, this was rarely observed during this study. Nevertheless, sample loss during melt transfer should not fractionate REEs from one another.

Horwitz et al. (2005) and Pourmand and Dauphas (2010) demonstrated that the *N,N,N'N' tetraoctyl-1,5-diglycolamide* ligand, commercially available as TODGA resin, has exceptionally high affinities for lanthanides and can be used for quantitative separation of the REEs from other matrix elements. Upon complete dissolution, all samples were passed through the TODGA resin and matrix elements were removed by eluting 12 mL of 3 mol $L^{-1}$ $HNO_3$ followed by 15 mL of 3 and 12 mol $L^{-1}$ $HNO_3$. Subsequently, the REEs were eluted in 30 mL of 0.05 mol $L^{-1}$ HCl (see Fig. 2 in Pourmand et al., 2012). The eluted fraction was evaporated at sub-boiling temperature under a heat lamp to a small droplet of about 1-5 µL inside a class-100 laminar flow hood and the residue was diluted in 2 mL of 0.45 mol $L^{-1}$ $HNO_3$ prior to analysis.

Measurements were made on Thermo Scientific Neptune Plus multi-collector ICP-MS instruments at the Origins Lab (the University of Chicago) and the Neptune Isotope Lab (the University of Miami – RSMAS). Additional details on the Neptune MC-ICPMS can be found in Wieser and Schwieters (2005). The analysis involved a novel dynamic Faraday cup configuration in which the zoom optics (focus and dispersion) and source lens (focus, X-Y deflection and shape) parameters of the Neptune instrument were adjusted to accommodate the wide mass range of 14 REEs ($^{139}$La-$^{175}$Lu) in 3 sub-configurations to measure all REEs in a single routine (Pourmand et al., 2012). To eliminate fluctuations in sensitivity between the 3 sub-configurations, $^{149}$Sm and $^{167}$Er measured in the middle REE configuration were also measured in LREE ($^{149}$Sm) and HREE ($^{167}$Er) sub-configurations.

Sample uptake was set at 90 s and data was acquired in dynamic mode through 1 block of 5 cycles, 4.2 s integration time and 3 s magnet settling (idle) time for each sub-routine. All samples were introduced into the plasma through desolvating nebulizers (ESI Apex-Q and Spiro) to minimize oxide interferences. The inlet system was flushed with 0.45 mol $L^{-1}$ $HNO_3$ for 120 s after each sample/standard solution measurement to eliminate memory effect from previous runs.

Each sample was measured up to three times and the REE abundances were quantified relative to a multi-element standard solution according to the relationship $C_A = C_S * (I_S / I_A)$, where $C_A$ and $C_S$ represent the concentrations in the sample and the multi-element standard, respectively, and $I_A$ and $I_S$ represent the intensities of the ion beams registered at the faraday detectors. As is often done in trace element studies, uncertainties are reported as the relative standard deviation (RSD) of the mean in percent [$100 \times (SD/\sqrt{n})/\bar{x}$, n=number of replicate analysis, $\bar{x}$=average concentration of separate digestions] unless otherwise specified. The REE patterns in this study are reported relative to the mean of CI-chondrites from Pourmand et al. (2012) updated with 2 additional measurements from this study (n=10 total; **Table 1**; La=0.24818, Ce=0.63662, Pr=0.09643, Nd=0.48805, Sm=0.15630, Eu=0.06004, Gd=0.21017, Tb=0.03785, Dy=0.25758, Ho=0.05510, Er=0.16539, Tm=0.02584, Yb=0.16860, Lu=0.02539 ppm).

REE anomalies represent deviations from neighboring elements based on the expected change in REE abundances as a function of atomic number and ionic radii (Lipin and McKay, 1989). Anomalies were calculated using linear interpolation of the logarithm of the CI-normalized abundances against ionic radius of neighboring elements. The anomalies were calculated according to the following equations:

$$Ce/Ce^* = Ce_N / (La_N^{0.48} \times Pr_N^{0.52}), \tag{1}$$

$$Eu/Eu^* = Eu_N / (Sm_N^{0.45} \times Gd_N^{0.55}), \tag{2}$$

$$Tm/Tm^* = Tm_N / (Er_N^{0.55} \times Yb_N^{0.45}). \tag{3}$$

In the text and figures, those anomalies are also reported as percent deviations from unity (*e.g.*, a +20 % Eu anomaly corresponds to Eu/Eu*=1.2). The following ionic radii are used for the 3+ 6-coordinate ions, La=103.2 pm, Ce=101.0, Pr=99.0, Nd=98.3, Pm=97; Sm=95.8, Eu=94.7, Gd=93.8, Tb=92.3, Dy=91.2, Ho=90.1, Er=89.0, Tm=88.0, Yb=86.8, Lu=86 (Shannon, 1976). Yb and Sm are relatively volatile under reducing conditions (Lodders and Fegley, 1993) and display anomalies of their own in enstatite meteorites (Hsu and Crozaz, 1998; Barrat et al., 2014). Yb is also more volatile than other HREEs in CAIs (Boynton, 1975; Davis and Grossman, 1979; Kornacki and Fegley, 1986). A potential concern is therefore that Eu and Tm anomalies could be affected by anomalies on the normalizing elements (Ireland and Fegley, 2000). For this reason, we also calculated those anomalies using Nd-Gd and Er-Lu for normalization (pronounce Eu and Tm double star),

$$\text{Eu/Eu}^{**} = \text{Eu}_N / \left( \text{Nd}_N^{0.2} \times \text{Gd}_N^{0.8} \right), \tag{4}$$

$$\text{Tm/Tm}^{**} = \text{Tm}_N / \left( \text{Er}_N^{0.66} \times \text{Lu}_N^{0.34} \right). \tag{5}$$

Another possible concern is that the curvature of the REE element pattern affected the Tm anomalies calculated based on simple linear interpolations. For this reason, we evaluated a third approach to normalizing Tm by using a Lagrangian (3$^{rd}$ order polynomial) interpolation between four REEs. In group II REE pattern, Dy, Ho, Er, and Lu form relatively smooth trends (Tanaka and Masuda, 1973, Martin and Mason, 1974, Grossman and Ganapathy, 1976; Mason and Taylor, 1982; Fegley and Ireland, 1991), so they were used in this third normalization (pronounce Tm triple star),

$$\text{Tm/Tm}^{***} = \text{Tm}_N / \left( \text{Dy}_N^{0.33} \times \text{Ho}_N^{-1.29} \times \text{Er}_N^{1.85} \times \text{Lu}_N^{0.11} \right) \tag{6}$$

In bulk meteorites, the Eu/Eu* and Eu/Eu** anomalies calculated are very similar. All three definitions of Tm anomalies (Eqs. 3, 5, and 6) have strengths and weaknesses. Tm/Tm* (Er-Yb normalization) leads to the most precise values because the normalizing REEs have relatively high abundances but it is not well suited for samples that could exhibit Yb anomalies such as enstatite meteorites or samples with strongly fractionated REE patterns such as terrestrial rocks. Tm/Tm** (Er-Lu normalization) is less precise because Lu is less abundant than Yb but it avoids the issue or normalizing to an element that can have anomalies. However, it does not address the issue of curvature in HREE pattern. Tm/Tm*** (Dy, Ho, Er, and Lu normalization) addresses the issue of Yb normalization and curvature in the REE pattern but the analytical precision is lower than for the other definitions. The Tm anomalies calculated using the three definitions agree for most samples. To avoid the issue of normalizing to Yb, we use the Tm/Tm** definition when discussing meteorite results. To avoid the issue of interpolating a curved REE pattern with a straight line, we use the Tm/Tm*** definition when discussing terrestrial results.

3. **Results**

The accuracy of our analytical technique was previously evaluated by processing replicates of geological reference materials BHVO-1 (basalt), BIR-1 (basalt), BCR-2 (basalt), PCC-1 (peridotite), G-2 (granite), G-3 (granite) and W-2 (diabase) from the USGS, which gave concentrations that were in agreement with certified and accepted values from the literature (Fig. 3 in Pourmand et al., 2012). The results for replicate analysis of G-3 CRM (n=6) and Orgueil primitive chondrite (n=7) from this study and Pourmand et al. (2012) are presented in **Table 1** and represent our best measure of external reproducibility. This is a conservative approach to assessing procedural reproducibility as natural samples can be heterogeneous and the concentrations of REEs in phosphate-bearing materials can vary due to the nugget effect (Sect. 4.1). Given the relatively small masses analyzed, the abundances of REEs in Orgueil may have been affected by sample heterogeneity. In comparison, the G-3 powder that was homogenized from ~ 272 kg of rock is more representative of the bulk composition, and yields better reproducibility; the standard deviations of Eu and Tm anomalies from replicates of G-3 are significantly smaller than replicates of Orgueil, presumably reflecting sample heterogeneity in the latter (**Table 1**). We therefore interpret Eu and Tm anomalies that exceed the uncertainties on G-3 to be potentially significant. Contributions from procedural blanks (n=9) were negligible for most REEs (**Table 1**). Nevertheless, the data were corrected by subtracting procedural blanks.

The concentrations of La-Lu in 41 chondrites from CI, CM, CO, CV, EH, EL, H, L, LL and two aubrites are presented in **Table 2**. Additional information on meteorite collection identification, recovery method (find *vs.* fall) and homogenized and digested masses are also provided along with Eu and Tm anomalies for each meteorite calculated according to equations 1 to 6. The samples represent 32 falls, 9 finds and duplicate analyses of the same or different specimens, including seven meteorites that were homogenized from large chips of the same collection identification. Previously, Dauphas and Pourmand (2011) measured the concentrations of Lu, Hf, U and Th in bulk samples of 37 meteorites by isotope dilution mass spectrometry. These measurements were mostly carried out on meteorite fragments or powder aliquots from the same meteorite specimens analyzed here. Lutetium concentrations by IDMS compare well with concentrations calculated by standard bracketing technique from this study (**Fig. 1**). A stronger agreement is achieved if St. Severin that shows the most discrepant concentration is not included. The reason for this discrepancy remains unknown. REE compositions of terrestrial samples measured using the same protocol as meteorites are compiled in Table 3. These are mostly published data (Pourmand et al., 2012, 2014) but they are included here to provide a direct comparison with extraterrestrial samples.

### 3.1. Carbonaceous chondrites

In addition to previously published results from Orgueil and Ivuna **(**Pourmand et al., 2012**),** which are included in **Table 2** for comparison, we present new data from three samples of Orgueil measured at the Neptune Isotope Lab; two new samples from a chip of MNHN (120) and one from a chip of NMNH 219 that was previously measured at the Origins Lab of the University of Chicago using the same analytical technique. This specimen had anomalously high abundances of light and middle REEs. As shown in **Fig. 2**, the anomalous abundance pattern for NMNH 219 is closely replicated. In contrast, the REE patterns for the new specimens of Orgueil appear flat and quite similar to previous measurements of Orgueil. Barrat et al. (2012) investigated the effect of sample size on the abundances of 46 elements in replicates of large Orgueil specimens using a Thermo Scientific Element 2 ICP-SFMS. Although these authors found varying degrees of heterogeneity, the REE concentrations in relatively large sample sizes of 0.5 to 1.2 g are indistinguishable from CI-chondrite values reported here and in Pourmand et al. (2012) on smaller masses (**Fig. 2**). Taken together, the agreement between Lu concentrations by isotope dilution and standard bracketing technique and the reproducibility of G-3 CRM and Orgueil measurements demonstrate the fidelity of our analytical technique.

In detail, differences between studies can be found, for instance in the geochemically important Sm/Nd ratio. Excluding one clear outlier (Orgueil H), the average Sm/Nd atomic ratio of 10 CI-chondrite measurements (1 Alais, 7 Orgueil, and 2 Ivuna) reported in **Tables 1 and 2** (this study; Pourmand et al., 2012) is 0.3072±0.0011 (95 % confidence interval of the mean). Bouvier et al. (2008) reported a value of 0.3121 based on isotope dilution analyses of one sample each of Orgueil and Ivuna. Barrat et al. (2012) reported a value of 0.3151±0.0029 based on analyses of 1 Alais, 1 Ivuna, and 6 Orgueil samples. The small (~1.5 %) but significant difference between those studies most likely reflects differences in absolute Sm/Nd ratio of the standard solutions used to calibrate the measurements. In this study, a multi-element standard solution with certified concentrations was used. The rationale for using a pre-mixed multi-element standard with certified composition is that the relative error on gravimetric determinations may be smaller when large batches of standard are prepared. The uncertainty on the absolute concentration of the Spex CertiPrep standard is ±0.5 %, corresponding to a relative uncertainty on elemental ratios of ±0.7 %, *i.e.*, ±0.22 on the Sm/Nd ratio. Even in single spike measurements (Bouvier et al., 2008), the spike concentrations are determined by counter-spiking a standard solution of known concentration. Absolute ratios are thus only as good as the standards and the small differences between studies may be due to differences in standardization. This is not an issue for the work discussed here as REE patterns are normalized to CI concentrations measured using the same standard.

The concentrations of REEs in three replicates of Murchison (CM2), which came from a large mass of homogenized powder (**Table 2**), show highly reproducible patterns characterized by CI-like abundances and a marked Gd anomaly (**Fig. 2**). Mighei (CM2) is enriched in REEs relative to CI-chondrites with a relatively flat abundance pattern (**Fig. 2**). Similarly, Lancé (CO3.5) and Kainsaz (CO3.2) are enriched relative to CI-chondrites and show relatively flat REE patterns (**Fig. 3**).

The distribution of REEs in Vigarano is characterized by depletion of elements of intermediate volatilities (La, Ce, Pr, Nd and Sm), enrichment in more refractory elements (Gd, Tb, Dy, Ho, Er and Lu) and negative anomalies in the most volatile elements, *i.e.,* Eu and Yb (**Fig. 3**). This is reminiscent of the pattern in ultra-refractory inclusions that are complementary to group II CAIs (McPherson et al., 1988). Grosnaja (CV3) shows enrichment in LREEs relative to HREEs and lacks a significant Eu or Tm anomaly (**Fig. 3**).

The abundance patterns for replicates of Allende (CV3) reference powder from the Smithsonian Institution (NMNH 3529, Split 8, position 5, Jarosewich et al., 1987) show relative enrichment in LREEs, depletion in HREEs, negative Eu anomaly and a prominent positive Tm anomaly (**Fig. 3**). This pattern has been extensively documented in measurements of bulk Allende (Nakamura, 1974, Jarosewich et al., 1987, Shinotsuka et al., 1995, Shinotsuka and Ebihara, 1997, Makishima and Nakamura, 2006, Pourmand et al. 2012, Barrat et al., 2012; Stracke et al., 2012) and is attributed to group II CAIs in carbonaceous chondrites (Martin and Mason, 1974; Mason and Taylor, 1982, Grossman and Ganapathy, 1976). The pattern is believed to result from differences in volatility between REEs during evaporation-condensation in the solar nebula (Boynton, 1975, Davis and Grossman, 1979, Kornacki and Fegley, 1986; MacPherson, 2004). Eu, and Yb are more volatile than LREEs (and Tm), which in turn are more volatile than ultra-refractory REEs Gd-Er and Lu. In group II REE pattern, the most refractory REEs (Gd-Er and Lu) are depleted presumably because they were removed in an ultra-refractory component that is missing from the CAIs and the most volatile REEs (Eu and Yb) are also depleted because they were never fully condensed. Tm stands out as an anomaly because unlike other HREEs Gd-Er and Lu that are ultra-refractory, it has intermediate volatility like LREEs. The greater volatility of Tm relative to other HREE is partly because gaseous Tm is mono-atomic while most other HREEs are present as LnO (Boynton, 1975 and Davis and Grossman, 1979). However, thermodynamic calculations fail to reproduce the large Tm anomalies measured in group II CAIs and further work is needed to understand quantitatively this pattern (Davis and Grossman, 1979; Kornacki and Fegley, 1986).

The Allende reference powder was homogenized from a large mass (4 kg) at the Smithsonian Institution (Jarosewich et al., 1987). Although the general REE pattern from replicates of this meteorite are consistent with group II-type CAIs (**Fig. 3**), the dispersions in Ce, Eu and Tm anomalies of replicate analyses exceed analytical uncertainties (**Table 1**). A compilation of these anomalies measured in various splits and positions of the reference powder by SSMS and NAA (Jarosewich et al., 1987), ICP-QMS (Shinotsuka et al., 1995 and Makishima and Nakamura, 1997), ICP-SFMS (Barrat et al., 2012), MC-ICPMS (Pourmand et al., 2012 and this study) is shown in **Fig. 4**. The small disagreements between different studies may be because different analytical methodologies were employed for samples from different splits and positions. An inter-laboratory study using a unified analytical methodology should reveal the extent of heterogeneity with respect to CAIs and other primary carriers of the REEs in the Smithsonian bulk Allende reference powder at the 50 mg scale due to incomplete homogenization or a nugget effect.

### 3.2. Enstatite chondrites and aubrites

Members of this group include unequilibrated and equilibrated EL (low Fe) and EH (high Fe) enstatite chondrites, and are known to have formed under highly reducing conditions (Keil, 1968; Larimer, 1968, 1975; Larimer and Bartholomay, 1979; Sears et al., 1982, Lodders and Fegley, 1993 and Grossman et al., 2008). The REEs in these meteorites are mainly associated with sulfide minerals such as oldhamite, and to a lesser degree with niningerite, alabandite and silicate minerals enstatite and plagioclase for Eu (Larimer and Ganapathy, 1987; Floss and Corzaz, 1993, Crozaz and Lundgerg, 1995, Hsu and Crozaz, 1998, Gannoun et al., 2011; Barrat et al., 2014).

The abundance patterns of the REEs in 9 EL chondrites with a replicate of Pillistfer (EL6), and six EH chondrites with a replicate of Indarch (EH4) are presented in **Fig. 5** and agree well with the recent study of Barrat et al. (2014). The bulk concentrations of REEs in most samples are depleted relative to CI-chondrites by ~0.6×CI. Notable exceptions are Eagle (EL6) and Hvittis (EL6) with slightly enriched HREEs, and Daniel's Kuil (EL6) and Blithfield (EL6), which are enriched in all REEs (**Fig. 5a**). The majority of ELs show a smooth trend of LREE depletion relative to HREE. The meteorites that are most enriched in REEs also have distinct features, notably more concave, downward REE patterns and prominent negative Eu anomalies. Happy Canyon (EL6/7) shows highly pronounced Ce and Eu anomalies and the highest enrichment in LREE among all enstatite chondrites analyzed in this study (not shown, **Table 2**). Happy Canyon was initially described as an enstatite achondrite (Olsen et al., 1977) but is now believed to be the product of impact-melting of EL chondrite material (Sears et al., 1982 and McCoy et al., 1995). The peculiar REE pattern of Happy Canyon most likely reflects terrestrial alteration as this sample is extensively weathered.

The abundances of REEs in EH chondrites are slightly lower than, or similar to CI-chondrites (**Fig 5b**). The patterns are flat overall, with some samples showing slight enrichments or depletions in HREEs relative to LREEs. The relatively unequilibrated Qingzhen (EH3) shows a slight enrichment in Gd-Er relative to CI.

Kallemeyn and Wasson (1986) argued that the LREE depletion in EL chondrites was caused by fractionation in the solar nebula. Floss and Crozaz (1993) also found several other REE patterns in oldhamite that were attributed to metamorphic processes, and thereby suggested that a combination of nebular and parent-body processes should be considered to explain the REE patterns and anomalies. In a recent study of equilibrated and unequilibrated EH and EL chondrites, Barrat et al. (2014) concluded that if EL6 chondrites are part of the same metamorphic sequence, a complementary reservoir is yet to be identified with LREE enrichments and positive Eu anomalies from which EL6 chondrites have evolved.

Aubrites are thought to be the melt products of an enstatite-chondrite precursor (Keil, 2010). The two aubrites analyzed are enriched in HREEs relative to LREEs and have marked positive (Bishopville) and negative (Norton County) Eu anomalies.

EH and EL chondrites display negative Tm anomalies (the average Tm/Tm*, Tm/Tm**, and Tm/Tm*** values are -3.0±0.6, -4.1±0.7, and -4.4±0.7 %, respectively) (Table 4). Happy Canyon is an outlier with a Tm/Tm* anomaly of -11 % but this value is sensitive to the normalization (Tm/Tm**=-6 % and Tm/Tm***=-5 %), suggesting that there may be a problem with Yb normalization (Tm/Tm*) for this particular sample.

### 3.3. Ordinary chondrites

The REE abundances of 14 ordinary chondrites (and 6 duplicates) from H, L and LL groups range from 1.1 to 6.5 × CI-chondrites (**Table 2**), and confirm that the parent-bodies of these chondrites have higher abundances of REEs relative to enstatite chondrites (Evensen et al., 1978). The majority of ordinary chondrites and their duplicate measurements from larger pieces fall into two complementary patterns; those in **Fig. 6a** are more depleted in LREE relative to HREE and show various degrees of positive Eu anomaly. Other samples either show flat patterns or are slightly enriched in LREE relative to HREE with various degrees of negative Eu anomalies (**Fig. 6b**). These distribution patterns are not specific to a particular meteorite group or petrologic type as L4, L6, LL4 and LL5 specimens appear in both clusters. The REE pattern in Bielokrynitschie (H4) is different from the rest of OCs with significant enrichment in La-Sm and a flat HREE pattern. This sample also carries the largest positive Ce anomaly (Ce/Ce* = +128.91%) among all meteorites from this study (**Table 2**).

All ordinary chondrites (H, L, and LL) display negative Tm anomalies. The mean Tm/Tm*, Tm/Tm** and Tm/Tm*** values of ordinary chondrites are -4.2±0.2, -4.2±0.3 and -4.8±0.4 % (Table 4). LL4 Hamlet is an outlier with a Tm/Tm* anomaly of ~-2.2 %.

### 3.4. Terrestrial rocks

The REE patterns of terrestrial rocks have been discussed extensively in the literature and will not be examined further here, except regarding Tm anomalies. Terrestrial rocks measured using the same analytical protocol as meteorites comprise a variety of rock types, including African dust, Post Archean Australian Shale (PAAS), and various geostandards (Table 4; this study; Pourmand et al., 2012, 2014). Terrestrial samples show negative Tm/Tm* anomalies similar to ordinary and enstatite chondrites. They also show significant dispersion in Tm/Tm* that correlates broadly with the degree of HREE fractionation as quantified by the $Dy_N/Lu_N$ ratio (Fig. 7). Terrestrial samples have experienced complex geological histories and can have highly fractionated REEs. The Tm/Tm* and Tm/Tm** use linear interpolations between Er-Yb and Er-Lu, respectively. To test whether curvature in the HREE pattern could explain the variations in Tm/Tm* and Tm/Tm** anomalies in terrestrial rocks, we calculated Tm/Tm** anomalies by normalizing to a Lagrangian (3$^{rd}$ order polynomial) fit of Dy, Ho, Er, and Lu abundances. Using this normalization, the dispersion in Tm anomalies collapses (including all data, the standard deviations of Tm/Tm*, Tm/Tm**, and Tm/Tm*** of 50 terrestrial rocks are 1.3, 1.1 , and 0.7 %) and the correlation between Tm anomalies and $Dy_N/Lu_N$ disappears (Fig. 7). We therefore recommend that the Tm/Tm*** notation be used for terrestrial samples with fractionated REE patterns.

### 4. Discussion

The REE patterns, $La_N/Lu_N$ ratios and Ce and Eu anomalies can vary among duplicate measurements of bulk meteorites (**Table 2**). The reason for these observed inconsistencies is most likely preferential concentration of REEs in certain minerals relative to the matrix such that samples homogenized from smaller chips may not be representative of the bulk composition (Jarosewich et al., 1987, Nittler et al. 2004, Morlok et al., 2006 and Barrat et al. 2012). Nevertheless, examining the REE patterns in detail can provide considerable information on the processes that governed elemental fractionation of refractory elements in the solar nebula and meteorite parent-bodies.

#### 4.1. Nugget effect during parent-body metamorphism

The influence of parent-body processes can be assessed by comparing the dispersion of redox-sensitive REE (Ce and Eu) anomalies and $La_N/Lu_N$ ratios (a proxy for fractionation of LREEs relative to HREEs) as a function of degrees of aqueous alteration or metamorphism (petrologic type). The most pristine meteorites are represented by petrologic type 3. The effect of aqueous alteration increases from 3 to 1 and thermal metamorphism prevails in higher petrologic types from 3 to 6. In planetary bodies, two different oxidation numbers in Ce and Eu result in different partitioning behaviors for these elements compared with other trivalent REEs. Aqueous alteration can also mobilize Ce in meteorites that were exposed to highly oxidizing terrestrial environments.

The Ce anomalies from this study show no correlation with petrologic type and some of the observed variations may have been produced by terrestrial weathering, especially in ordinary and more reduced enstatite chondrites, which are prone to oxidative weathering (Floss and Crozaz, 1991, Crozaz and Wadhwa, 2001, Crozaz et al., 2003). In contrast to Ce anomalies, the dispersions of $La_N/Lu_N$ ratios and Eu anomalies increase as function of petrologic type 3 to 6 and are in agreement with a strong influence of thermal metamorphism (**Fig. 8a and b**). A similar relationship was reported in the dispersion of $^{176}Hf/^{177}Hf$, Lu/Hf, and Th/Hf ratios as a function of the degree of metamorphism (Dauphas and Pourmand 2011). Bouvier et al. (2008) and Martin et al. (2013) also noted that $^{176}Hf/^{177}Hf$ and Lu/Hf ratios were much less variable in unequilibrated chondrites compared to meteorites that had experienced some metamorphism.

Nakamura (1974) found correlations between REE abundances and Eu anomalies in chondrites and similar relationships are found here (**Fig. 9**). As discussed above, the variations in Eu anomalies in metamorphosed chondrites most likely have a parent-body origin, arising from REE redistribution in phosphates or sulfides. We can therefore filter for the effect of metamorphism on ordinary and enstatite parent-bodies excluding those samples that show pronounced Eu anomalies. In **Fig. 10**, we report the REE patterns of all ordinary and enstatite chondrites that have no clearly resolvable Eu anomalies (within -5 and +5 %). Ordinary and enstatite chondrites filtered by this criterion have relatively flat REE patterns, strengthening the relevance of CI chondrite REE abundances as reference values in solar system studies. As discussed below, thulium is an exception as ordinary and enstatite chondrites display anomalies for this element relative to CI.

At first sight, the finding that metamorphism in chondrites tends to increase the scatter of REE ratios and concentrations may be counterintuitive as metamorphism tends to erase grain-to-grain chemical and isotopic heterogeneities that exist in unequilibrated chondrites (Van Schmus and Wood, 1967; Grossman and Brearley, 2005). To understand what causes REE dispersion in metamorphosed chondrites, it is important to know what the main carrier phases are. In enstatite chondrites, REEs are carried by sulfide minerals (oldhamite), enstatite and plagioclase for Eu (Larimer and Ganapathy, 1987; Ebihara, 1988; Crozaz and Lundberg, 1995; Hsu and Crozaz, 1998; Gannoun et al., 2011b; Barrat et al., 2014). In ordinary chondrites of low metamorphic grade, REEs are carried by glass in chondrule mesostatis (Alexander, 1994). As metamorphism increases and chondrule mesostasis crystallizes, the REEs partition into Ca-phosphate minerals, Ca-pyroxene, and plagioclase for Eu (Curtis and Schmitt, 1979; Ebihara and Honda, 1983, 1984; Murrell and Burnett, 1983; Crozaz and Zinner, 1985; Crozaz et al., 1989; Alexander, 1994). In metamorphosed chondrites, most of the LREEs and half of the LREEs are in phosphate (the other main carrier phase is pyroxene). Europium is concentrated in plagioclase. The phosphates are thus depleted in HREEs relative to LREEs and have marked negative Eu anomalies. We interpret the dispersion in REE patterns and Eu anomalies in metamorphosed ordinary chondrites to reflect the nugget effect arising from the concentration of REEs in minor phosphate grains and the relatively small mass of sample digested in most studies. Below we develop a quantitative model to explain this phenomenon.

The mass of meteorite sampled for measurement is noted $m$. The average densities of the nugget mineral, the matrix (*i.e.*, all non-nugget material), and the bulk are $\rho_{\text{nugget}}$, $\rho_{\text{matrix}}$, and $\rho_{\text{bulk}}$. The mineral responsible for the nugget effect (phosphate here) represents a volume fraction $f$ of the total sample. One can think as individual mineral grains as imbricated bricks that are stacked in three dimensions. For simplicity, we shall assume that all grains in the sample have the same size (equivalent diameter $d$) as the nugget mineral and there are $N$ such grains in the sample mass $m$ ($\lfloor x \rfloor$ in the nearest integer to $x$),

$$N = \frac{\text{Sample volume}}{\text{Individual nugget volume}} = \left\lfloor \frac{6m}{\pi d^3 \rho_{\text{bulk}}} \right\rfloor$$
(7)

Based on this simple description, one can calculate the probability that the sample mass contains $n = k$ nugget grains, which we note $P(n = k)$. The probability for each individual grain to be a nugget mineral is $f$. $P(n = k)$ is therefore given by the Binomial distribution,

$$P(n = k) = \binom{N}{k} f^k (1 - f)^{N-k}. \tag{8}$$

The mean and standard deviation of this distribution are $\bar{n} = Nf$ and $\sigma_n = \sqrt{Nf(1-f)}$. A Poisson distribution of mean $Nf$ is a fine approximation, a well-known result when evaluating nugget occurrences in economic geology (Clifton *et al.*, 1969),

$$P(n = k) \simeq (Nf)^k e^{-Nf}/k! \tag{9}$$

If the nugget mineral contains an element at concentration $C_{\text{nugget}}$ and the rest of the sample has a concentration $C_{\text{matrix}}$ (both in g/g), $k$ nuggets in the sample mass $m$ corresponds to a bulk concentration,

$$C_{\text{bulk}} = \frac{k\rho_{\text{nugget}}C_{\text{nugget}}+(N-k)\rho_{\text{matrix}}C_{\text{matrix}}}{k\rho_{\text{nugget}}+(N-k)\rho_{\text{matrix}}} \tag{10}$$

The bulk concentration follows approximately a normal distribution of mean and standard deviation (these equations were obtained by replacing $Nf$ for $k$ and propagating the uncertainty on $k$):

$$\bar{C} = \frac{f\rho_{\text{nugget}}C_{\text{nugget}}+(1-f)\rho_{\text{matrix}}C_{\text{matrix}}}{f\rho_{\text{nugget}}+(1-f)\rho_{\text{matrix}}}, \tag{11}$$

$$\sigma_C = \frac{|C_{\text{nugget}}-C_{\text{matrix}}|\rho_{\text{nugget}}\rho_{\text{matrix}}}{[(1-f)\rho_{\text{matrix}}+f\rho_{\text{nugget}}]^{3/2}} \sqrt{\frac{f(1-f)\pi d^3}{6m}}. \tag{12}$$

Given that $f$ is small, those two equations can be approximated by,

$$\bar{C} \simeq C_{\text{matrix}} + f(\rho_{\text{nugget}}/\rho_{\text{matrix}})C_{\text{nugget}}, \tag{13}$$

$$\sigma_C \simeq |C_{\text{nugget}} - C_{\text{matrix}}|(\rho_{\text{nugget}}/\rho_{\text{matrix}})\sqrt{\frac{f\rho_{\text{matrix}}\pi d^3}{6m}}. \tag{14}$$

Considering 2 elements of concentrations $C_1$ and $C_2$, the ratio $R = C_2/C_1$ in the bulk will approximately follow a normal distribution of mean and standard deviation (we note $r = C_{1,\text{nugget}}/C_{1,\text{matrix}}$):

$$\bar{R} = \frac{f\rho_{\text{nugget}}R_{\text{nugget}}r+(1-f)\rho_{\text{matrix}}R_{\text{matrix}}}{f\rho_{\text{nugget}}r+(1-f)\rho_{\text{matrix}}}, \tag{15}$$

$$\sigma_R = \frac{r|R_{\text{nugget}} - R_{\text{matrix}}|\rho_{\text{nugget}}\rho_{\text{matrix}}}{[(1-f)\rho_{\text{matrix}} + rf\rho_{\text{nugget}}]^2} \sqrt{\frac{f(1-f)\pi d^3[(1-f)\rho_{\text{matrix}} + f\rho_{\text{nugget}}]}{6m}}. \tag{16}$$

Given that $f$ is small, those two equations can be approximated by,

$$\bar{R} \simeq R_{\text{matrix}} + \frac{rf(\rho_{\text{nugget}}/\rho_{\text{matrix}})}{rf(\rho_{\text{nugget}}/\rho_{\text{matrix}}) + 1}(R_{\text{nugget}} - R_{\text{matrix}}), \tag{17}$$

$$\sigma_R \simeq \frac{r(\rho_{\text{nugget}}/\rho_{\text{matrix}})}{[1 + rf(\rho_{\text{nugget}}/\rho_{\text{matrix}})]^2} \sqrt{\frac{f\rho_{\text{matrix}}\pi d^3}{6m}} |R_{\text{nugget}} - R_{\text{matrix}}|. \tag{18}$$

The masses of ordinary chondrites digested ranged between 36 and 160 mg, with an average value of 85 mg. These digestions were prepared from homogenized powders prepared from samples weighing 110 to 940 mg, with an average value of 450 mg. Powdering ordinary chondrites is difficult because they contain metal grains than cannot be crushed in an agate mortar, thereby limiting the efficiency with which other grains, such as phosphates, can be crushed. For this reason, we take $m = 85$ mg as representative. The standard deviation of elemental ratios scales as the square root of the mass homogenized (Eq. 18), which varies from sample to sample. For example, using the average homogenized mass of $m = 450$ mg would lead to a predicted dispersion that is 2.3 times smaller than that calculated using $m = 80$ mg.

Dreeland and Jones (2011) studied the size distribution of phosphate grains in metamorphosed ordinary chondrites. They found that the mode in the size distribution was between 10-50 μm but the distribution is skewed towards higher values and a significant number of grains are larger than 200 μm. Another consideration that has to be taken into account is the fact that large grains carry more mass than smaller grains because the volume scales as the cube of the radius. Taking this into account, we calculate weighted average merrillite grain sizes for Bjurböle (L/LL4), Tuxtuac (LL5), and St Séverin (LL6) of ~161, 196, and 193 μm, respectively. The weighted average grain sizes of chlorapatite for the same meteorites are 212, 184, and 151 μm, respectively. Therefore, we take a grain size (diameter) of 200 μm as representative of phosphates in metamorphosed ordinary chondrites. The volume fraction of phosphate minerals in metamorphosed ordinary chondrites is ~0.45 % ($f = 0.0045$) (Crozaz et al., 1989; Jones *et al.*, 2014). Those phosphates comprise apatite (0.11 %) and merrillite (0.34 %) (Jones *et al.*, 2014). The density of phosphate is 3.2 g/cm³, which is close to the density of the matrix or silicate minerals, so we consider a constant density $\rho = 3.2$ g cm$^{-3}$. The REE and Hf concentrations in phosphate and non-phosphate phases are compiled in **Table 5** (Wasson and Kallemeyn, 1988; Crozaz et al., 1989; Martin et al., 2013).

In **Fig. 11**, we plot the (La/Lu)$_N$, Eu/Eu*, $^{176}$Hf/$^{177}$Hf variations measured in metamorphosed ordinary chondrites (types 4-6) in the form of histograms (normalized as probability density functions) and a scatter-plot of (La/Lu)$_N$ *vs*. Eu/Eu*. These observations are compared with probability density functions calculated using the equations derived above (Eqs. 9, 10, 17, 18) assuming that a nugget effect associated with the concentration of REEs in phosphates is the cause for this dispersion. For reference, an 85 mg chondrite sample would only contain ~28 phosphate grains of 200 μm size and the 95 % confidence interval on this value is between 18 and 39 grains. As shown, the theoretical expectations match well the observations, indicating that a phosphate nugget effect is indeed responsible for much of the variations in REE patterns and abundances measured in ordinary chondrites. Data is missing to do the same calculation for enstatite chondrites but it is reasonable to assume that oldhamite plays the same role in these meteorites as phosphate in ordinary chondrites.

### 4.2. Nebular fractionation and thulium anomalies

The Tm anomalies measured in meteorite and terrestrial samples from this study, Pourmand et al. (2012) and Pourmand et al. (2014) are reported in **Tables 2**, 3 and 4 and displayed in **Fig. 8**c and **Fig. 12**. Allende (CV3) exhibits large, positive Tm anomalies that are associated with variations in other REEs and are reminiscent of group II REE patterns found mostly in fine-grain refractory inclusions (**Fig. 3**; Tanaka and Masuda, 1973, Martin and Mason, 1974, Grossman and Ganapathy, 1976; Mason and Taylor, 1982; Fegley and Ireland, 1991). Other carbonaceous chondrites do not have detectable Tm anomalies relative to CI chondrites.

Terrestrial rocks, ordinary and enstatite chondrites, and aubrites predominantly show negative Tm anomalies relative to the mean of CI-chondrites of ~-4.5 % (Table 4). Mono-isotopic Tm is not commonly measured in bulk samples; including in the most recent high-precision REE dataset of 26 chondrites from Barrat et al. (2014), so we cannot compare our data with literature values of similar precision. Bendel et al. (2012) and Bendel (2013; PhD thesis) found Tm anomalies in terrestrial and extraterrestrial samples that agree with the results presented here.

In contrast to $La_N/Lu_N$ and Eu anomalies that seem to be influenced by the petrologic grade of the meteorites considered, negative Tm anomalies in ordinary and enstatite chondrites are consistently reproduced (**Fig. 8c** and **Fig. 12**). Among REEs, Tm has intermediate volatility and behaves like LREEs during evaporation/condensation. Specifically, Kornacki and Fegley (1986) showed that Tm falls between Th and Nd in the volatility sequence derived from CAI analyses and thermodynamic calculations. Tm is more volatile than neighbor Er and is more refractory than neighbor Yb. Therefore, Tm anomalies most likely reflect nebular processes, as high temperature evaporation/condensation is the only natural setting known to fractionate Tm relative to neighbor lanthanides in a way that does not follow the general trends of abundance *vs*. ionic radius**.** Similar anomalies of larger magnitude have been identified in CAIs and are a characteristic of group II REE patterns (Tanaka and Masuda, 1973, Martin and Mason, 1974; Grossman and Ganapathy, 1976; Mason and Taylor, 1982). In Group II REE pattern, LREEs and Tm are present in their full complements and appear to be enriched relative to other REEs. This reflects sequestration of an ultrarefractory component (possibly carried by hibonite, MacPherson and Davis, 1994) in the condensation sequence prior to complete condensation of volatile elements (Boynton, 1975; Davis and Grossman, 1979).

In **Fig. 13**, we calculate synthetic REE patterns by adding or subtracting fine grain CAI material with group II REE patterns (Mason and Taylor, 1982). When only 0.3 % of group II CAI matter is added, this is enough to create a +10 % Tm anomaly similar to what is measured in Allende (Pourmand et al. 2012, Barrat et al., 2012; Stracke et al., 2012). The main collateral effect is a negative Eu anomaly, which is also seen in the bulk Allende REE pattern (**Fig. 3**). The difference between the synthetic and observed pattern for Allende is that MREEs (Sm, Eu and Gd) are not as depleted in the meteorite as expected for the mixing calculation, so components other than group II CAIs may also be involved. When removing 0.15 % of group II CAI to CI, one produces a -4.5 % Tm anomaly, similar to what is seen in ordinary, enstatite chondrites, and terrestrial rocks. Overall, the REE pattern predicted from the mixing calculation is relatively smooth except for the presence of a potentially resolvable Eu anomaly of +3.3 %. Much larger Eu variations are found in ordinary and enstatite chondrites (from -40 to +20 %, see **Table 2**) but these are primarily due to a nugget effect associated with redistribution of REEs during parent-body metamorphism (**Figs. 8b, 11**; Sect. 4.1). This metamorphic overprint prevents us from identifying the more subtle variation in Eu anomaly produced by condensation/evaporation in the solar nebula. By focusing in the future on ordinary and enstatite chondrites of low metamorphic grades, it may be possible to detect the predicted +3 % Eu anomaly if the Tm depletion indeed reflects a missing group II REE contribution.

Ordinary and enstatite chondrites as well as terrestrial rocks have fairly homogeneous Tm anomalies of around –4 to -6 % relative to CI chondrites. Other groups of carbonaceous chondrites show no significant (CM, CO, and CV except Allende) or large positive (Allende) Tm anomalies (**Fig. 12**). Because of the close match in abundances between CI chondrites and solar photosphere data for non-atmophile elements, CI chondrites are usually taken to represent the composition of the solar system as a whole (Anders and Grevesse, 1989, Palme and Jones, 2003, Lodders et al., 2009; Palme et al., 2014). There is no way of telling whether this is correct for lanthanides but assuming that it is, it should follow that most inner solar system bodies are depleted in a highly refractory component that hosted group II REE pattern with positive anomalies in Tm. This should therefore have created mirror enrichment in the residual gas from which terrestrial planets formed that is largely missing from meteorite collections, except for Allende. Alternatively, the compositions of Earth, ordinary and enstatite chondrites are representative of the bulk solar system value and CI chondrites are slightly enriched in a refractory component. At present, there is no way to tell which interpretation is correct. In Table 6, we calculate the composition if 0.15 % of a fine grained CAI component with group II REE pattern was removed from CI composition (Palme et al., 2014). This revised CI composition is noted CI* and may be better suited to normalize the compositions of inner solar system bodies that display Tm anomalies. The collateral effects on refractory element abundances of removing such a group II component are small and would be undetectable by solar spectroscopy. Further work is needed to characterize the chemical composition of group II CAIs to better define the composition of the CI* reference.

The negative Tm anomalies in ordinary and enstatite chondrites and the Earth relative to carbonaceous chondrites indicate that carbonaceous chondrites likely represent a small fraction of the material that made the terrestrial planets. This agrees with the conclusion based on isotopic anomalies in planetary bodies for $^{17}$O, $^{48}$Ca, $^{50}$Ti, $^{54}$Cr, $^{62}$Ni, and $^{92}$Mo that carbonaceous chondrites must represent a small fraction of the material that made the terrestrial planets (probably less than ~10 %; Lodders and Fegley, 1997; Lodders, 2000; Warren, 2011 and Dauphas et al., 2014a,b).

Huang et al. (2012) measured calcium isotopic ratios and REE abundances in refractory inclusions from the Allende CV3 chondrite. They found a broad correlation between Tm anomalies and Ca isotopic fractionation that they ascribed to evaporation/condensation processes in the solar nebula. Simon and DePaolo (2010), Huang and Jacobsen (2012), and Valdes et al. (2014) measured the isotopic composition of Ca in bulk chondrites. Although there are discrepancies between these three studies, they all agree that significant differences are present between chondrite groups. In particular, carbonaceous chondrites tend to have lighter Ca isotopic composition relative to ordinary and enstatite chondrites. This is reminiscent of what we have found for Tm and it is possible that the broad correlation found in refractory inclusions between Tm anomalies and $\delta^{44/40}$Ca is also present at the bulk meteorite scale.

Because carbonaceous chondrites show some heterogeneity at the scale of the hand-specimen for REEs and few samples have been analyzed for both Tm anomalies and Ca isotopic fractionation, comparing our results with the above-mentioned studies is not straightforward. In **Fig. 14**, we show the relation between Tm anomalies and Ca isotopic fractionation for the few samples for which this comparison can be made. Also plotted are mixing curves between the bulk silicate Earth (BSE) and the 4 group II refractory inclusions measured by Huang et al. (2012). Indeed, there may be a broad correlation between Tm anomalies and Ca isotopic fractionation that will take further data to fully resolve (*i.e.*, by measuring Tm anomalies and Ca isotopic fractionation on the same meteorite fragments). Some of the data points can be explained well by admixtures of type-II CAI-like material to the BSE composition. The fractionation of REEs from one another to create type-II REE patterns is thought to have taken place at a specific temperature window corresponding to the condensation temperature of hibonite (Beckett and Stolper, 1994, MacPherson and Davis, 1994) or perovskite (Boynton, 1975; Davis and Grossman, 1979). The stoichiometry of these phases is such (Ca/Al = 1/12 for hibonite and Ca/Ti = 1 for perovskite) that a relatively small fraction of total Ca is expected to have condensed by the time the group II REE pattern is established (Yoneda and Grossman, 1995). Ca condenses at lower temperature into melilite, so it can be only taken as an indirect proxy for the behavior of REEs during evaporation and condensation. Titanium condenses into perovskite (and possibly other calcium titanates; Lodders, 2002) at a temperature similar to the REEs and is expected to be a better isotopic proxy, so a better correlation is expected between REEs and Ti isotope systematics.

## 5. Conclusions

The bulk rare earth element compositions of 41 chondrites (**Table 2**) were analyzed in dynamic mode using a multi-collector inductively coupled plasma mass spectrometer, allowing us to measure the relative abundance of mono-isotopic REEs with high precision. Increased dispersion in $La_N/Lu_N$ ratio (Fig. 8a), Eu anomalies (Fig. 8b), and $^{176}Hf/^{177}Hf$ ratio with enhanced metamorphism is indicative of parent-body redistribution of REEs during metamorphism in chondrites. A model is presented to predict variations in elemental and isotopic ratios resulting from a nugget effect (Eq. 18). The high concentrations of REEs in phosphates in ordinary chondrites and the limited sample masses digested can explain quantitatively much of the variations in Eu anomalies, LREE/HREE, and $^{176}Hf/^{177}Hf$ ratio in metamorphosed ordinary chondrites (**Fig. 11**). Data is missing to do the same calculation for enstatite chondrites but it is reasonable to assume that oldhamite plays the same role in these meteorites as phosphate in ordinary chondrites.

As previously documented by others, some of the anomalous patterns previously identified in refractory inclusions as type-II REE patterns are also present in the bulk Allende carbonaceous chondrite sample, reflecting solid-gas fractionation processes in the solar nebula (**Fig. 3**). Addition of 0.35 % of fine-grain CAI with group II REE pattern to CI reproduces approximately the REE pattern measured in the Allende meteorite (**Fig. 13**).

A more subtle difference is found for mono-isotopic thulium in terrestrial rocks and ordinary and enstatite chondrites, which all show negative anomalies relative to CI chondrites of about -5 % (**Fig. 12**). This may reflect the fact that Earth and ordinary and enstatite chondrites formed from a reservoir of nebular gas from which refractory dust characterized by positive Tm anomalies had been sequestered. Removal of 0.15 % group II CAI from CI can explain the observed deficit in Tm (**Fig. 13**). Conversely, CI chondrites may contain an extra amount of a component akin to group II CAI relative to the composition of the Sun and inner solar system objects. A composition corrected for the presence of this component is calculated (CI*, Table 6) that may be better suited for normalizing the compositions of inner solar system objects. Carbonaceous chondrites must have represented a minor fraction of the constituents of the Earth, confirming previous inferences from isotopic anomalies.

Tm anomalies may be correlated with calcium isotopic fractionation in chondrites and other planetary bodies, reflecting fractionation of refractory elements by evaporation/condensation in the solar nebula (**Fig. 14**). A better correlation is expected between Tm anomalies and Ti isotopic mass fractionation, as fractionation of REEs and condensation of Ti are thought to have happened contemporaneously.

**Acknowledgements:** We thank the Field Museum, the Muséum National d'Histoire Naturelle, American Museum of Natural History and Natural History Museum in Vienna for proving meteorite samples. We are also grateful to R. N. Clayton for a sample of Qingzhen (EH3) and Thomas J. Ireland, Arash Sharifi and Junjun Zheng for their assistance with purifying the commercial lithium metaborate flux and mass spectrometry. We are grateful to T.R. Ireland, M. Norman, and an anonymous reviewer for their comments that greatly improved the manuscript. Discussions with J.-A. Barrat,  A.M. Davis, and L. Grossman, and were greatly appreciated. J.-A. Barrat indicated to us that Tm/Tm* anomalies in terrestrial rocks are correlated with REE fractionation. F.L.H. Tissot provided mean REE concentrations of fine grain CAIs with group II REE patterns. This work was funded by grants from NASA, (NNX12AH60G and NNX14AK09G) and NSF (EAR1144429 and EAR1444951) to ND and NSF (EAR-1003639) to AP.

**Figure captions**

**Fig. 1.** The concentrations of Lu measured mostly on different aliquots of chondrites by standard-sample bracketing technique (this study) agree well with measurements by isotope dilution mass spectrometry (Dauphas and Pourmand, 2011). St. Severin appears to be an outlier but the reason for the discrepancy for this meteorite is unknown.

**Fig. 2.** CI and CM REE abundances (**Table 2**) normalized to the CI composition from Pourmand et al. (2012) updated with the 2 new flat CI compositions measured in this study. The same CI composition is used for normalization in the following figures. Three chips of Orgueil are compared with literature measurements (Pourmand et al. 2012 and Barrat et al., 2012). Enrichment of LREEs in Orgueil MNHN-219 reported in Pourmand et al. (2012) is reproduced while two other samples from different chips show flat patterns relative to the mean of CI-chondrites. Mighei and Murchison show relatively flat REE patterns but Murchison has a marked negative Gd anomaly.

**Fig. 3.** CI-normalized REE abundances in CO and CV chondrites (**Table 2**). Replicates of NMNH 3529 Allende (CV3) reference material (split 8, position 5, Jarosewich et al., 1987) from this study and Pourmand et al. (2012) are similar to group II-type refractory inclusions (MacPherson et al., 1988). The pattern in Vigarano (CV3) is reminiscent of ultra-refractory inclusions. Grosnaja (CV3) is enriched in LREEs and shows flat HREEs with the exception of depletion in Lu. The two CO meteorites (Kainsaz and Lancé) have approximately flat REE patterns. All CO and CV chondrites are enriched in REEs relative to CI-chondrites.

**Fig. 4.** Ce, Eu and Tm anomalies in replicate measurements of NMNH 3529 Allende (CV3) reference material (Jarosewich et al., 1987) from this study (**Table 2**) are compared with the literature (Jarosewich et al., 1987, Shinotsuka et al., 1995, Makishima and Nakamura, 1997, Pourmand et al., 2012 and Barrat et al., 2012). The largest dispersions are observed in neutron activation analysis (NAA) and spark source mass spectrometry (SSMS), although smaller discrepancies persist among ICP-MS measurements of the same splits and positions. S/P represents different split/position of the reference material.

**Fig. 5.** CI-normalized REE patterns in a) EL and b) EH enstatite chondrites (**Table 2**). Except for Ilafegh 009 (EL7), all EL chondrites are from petrologic type six and most show a smooth trend of LREEs depletion relative to HREEs. Both positive and negative Eu anomalies are observed in EL chondrites. The REE patterns in EH chondrites are mostly flat with various

degrees of enrichment or depletion in HREEs. Various degrees of positive Ce and Tm anomalies are present in enstatite chondrites.

**Fig 6.** CI-normalized REE abundances in ordinary chondrites (**Table 2**) fall into complementary grouplets of a) depletion in LREEs and positive Eu anomaly, and b) Enrichment in HREEs or a flat pattern with negative Eu anomaly. Significant positive and negative Ce and Tm anomalies, respectively, are present in most ordinary chondrites.

**Fig. 7.** Relationships between Tm anomalies and $Dy_N/Lu_N$ ratio (a proxy for HREE fractionation) in terrestrial rocks (Table 3). A broad correlation is observed between Tm/Tm* anomaly (defined using Er and Yb; Eq. 3) and HREE fractionation. Normalizing to a Lagrangian fit to Dy, Ho Er, and Lu (Tm/Tm*** notation) removes this correlation, suggesting that it is an artifact arising from using a linear interpolation when the HREE pattern is strongly fractionated and shows some curvature.

**Fig. 8.** Dispersions of a) $La_N/Lu_N$, b) Eu and c) Tm/Tm** anomalies in chondrites as a function of petrologic type. The gray vertical bars represent the weighted mean of each group. The dispersion of $La_N/Lu_N$ and Eu anomalies increases as a function of metamorphism (petrologic types 3 to 6) on the parent bodies. The dispersion of Tm anomalies does not change with increased degrees of equilibration. The results for Allende (CV3) reference material, which include measurements from Pourmand et al. (2012) and Barrat et al. (2014), reveal a significant positive Tm anomaly.

**Fig. 9.** Correlations between LREE/HREE fractionation ($La_N/Lu_N$; left panels), lanthanide abundances (average enrichment factors of LREE La-Sm abundances normalized to CI; right panels), and Eu anomalies in carbonaceous (top panels), enstatite (middle panels), and ordinary (bottom panels) chondrites. These variations are due in part to nebular fractionation in CV and other carbonaceous chondrites and parent-body metamorphism in enstatite and ordinary chondrites (see text for details).

**Fig. 10.** REE patterns of chondrites with Eu anomalies within 5 % of the CI value. All samples have relatively flat REE patterns with no large anomaly, except for Tm (see text for details). Note that one outlier (Bielokrynitschie, H4) was not included in this figure.

**Fig. 11**. Comparison between measured and predicted dispersion in REE patterns in metamorphosed ordinary chondrites (types 4-6) due to phosphate nugget effect (Eqs. 9 to 18). The histograms (A, B, C; normalized as probability density distributions –PDF) represent the variations measured in metamorphosed ordinary chondrites (this study, Dauphas and Pourmand, 2011). Panel D shows a correlation (**Fig. 9**) between (La/Lu)N ratio (a proxy of LREE/HREE fractionation) and Eu anomalies. In all panels, the red curves show the expectations for a nugget effect associated with the concentration of REEs in phosphates that have fractionated REE patterns (high LREE/HREE and large negative Eu anomalies). Those curves were calculated using Eqs. 9, 10 and 18 and the data in **Table 5**. The red curve in panel D corresponds to the 95% confidence interval of the phosphate/non-phosphate mixing curve. As shown, phosphate nugget effect and limited sample mass in ordinary chondrites measurements are the causes for much of the REE fractionation in macroscopic ordinary chondrite measurements.

**Fig 12.** Tm anomalies among terrestrial rocks and meteorites (Tables 2 and 3). The weighted mean of meteorite groups and terrestrial material are shown at 95% confidence interval of the mean. Ordinary and enstatite chondrites, aubrites, and terrestrial rocks show systematic negative Tm anomalies of about -4.5 % relative to carbonaceous chondrites. Terrestrial rocks include replicates of reference materials and Post-Archaean Australian Shale (Pourmand et al., 2012) and trans-Atlantic African dust (Pourmand et al., 2014). See text for details on calculation of Tm anomalies.

**Fig. 13.** Synthetic REE patterns calculated by mixing CI composition (Pourmand and Dauphas, 2012) with group II REE patterns commonly found in fine grain inclusions (Mason and Taylor, 1982). The top curve corresponds to addition of 0.35 % group II CAI, which produces a +10 % Tm anomaly and reproduces approximately the REE pattern measured in Allende; **Fig. 3**). The bottom curve corresponds to removal of 0.15 % of group II CAI, which produces a -4.5 % Tm anomaly, which corresponds approximately to the anomalies found in ordinary, enstatite chondrites, and terrestrial rocks (**Fig. 12**). A collateral anomaly is predicted for Eu that is seen in the Allende pattern (**Fig. 3**) but is completely overprinted in ordinary and enstatite chondrites by metamorphism, which redistributed and fractionated REEs (**Fig. 11**).

**Fig. 14**. Relation between Tm anomalies (**Table 2**) and Ca isotopic fractionation (Valdes et al., 2013) in bulk meteorites. The black lines are mixing curves between BSE and type-II refractory inclusions analyzed by Huang et al. (2012). The labels on the curves correspond to the names of the refractory inclusions used in mixing. If carbonaceous chondrites are enriched in such refractory material relative to non-carbonaceous chondrites (or vice-versa if non-

carbonaceous chondrites are depleted), this could explain the dispersion in Tm anomalies and Ca isotopic fractionation among bulk meteorites.

**Table**

Table 1. Rare earth element concentrations (µg g$^{-1}$) in replicates of G-3, Orgueil, procedural blanks and the mean of CI-chondrites from this study and Pourmand et al. (2012).

| Element | G-3 (n=6) | RSD$_{mean}$ (%) | Orgueil (n=7) | RSD$_{mean}$ (%) | CI-Chondrite Mean (n=10)[1] | RSD$_{mean}$ (%) | Procedural Blank (pg, n=9) |
|---|---|---|---|---|---|---|---|
| La | 87.88 | 0.93 | 0.2481 | 0.61 | 0.2482 | 0.42 | 163 |
| Ce | 163.7 | 0.89 | 0.6372 | 0.66 | 0.6366 | 0.46 | 219 |
| Pr | 16.42 | 0.86 | 0.0966 | 0.47 | 0.0964 | 0.34 | 31 |
| Nd | 53.28 | 0.81 | 0.4882 | 0.67 | 0.4880 | 0.48 | 153 |
| Sm | 7.005 | 0.74 | 0.1563 | 0.66 | 0.1563 | 0.47 | 35 |
| Eu | 1.305 | 0.86 | 0.0601 | 0.56 | 0.0600 | 0.42 | 12 |
| Gd | 4.407 | 0.79 | 0.2099 | 0.74 | 0.2102 | 0.57 | 48 |
| Tb | 0.4520 | 0.87 | 0.0379 | 0.50 | 0.0379 | 0.40 | 10 |
| Dy | 2.056 | 0.86 | 0.2571 | 0.79 | 0.2576 | 0.64 | 40 |
| Ho | 0.3380 | 0.86 | 0.0550 | 0.92 | 0.0551 | 0.70 | 10 |
| Er | 0.8588 | 0.84 | 0.1645 | 1.2 | 0.1654 | 0.92 | 23 |
| Tm | 0.1138 | 0.92 | 0.0256 | 1.2 | 0.0258 | 0.93 | 5 |
| Yb | 0.7173 | 0.89 | 0.1675 | 1.1 | 0.1686 | 0.88 | 20 |
| Lu | 0.1007 | 0.90 | 0.0249 | 1.2 | 0.0254 | 1.5 | 6 |
| La$_N$/Lu$_N$ | 89.26 | 0.54 | 1.0204 | 1.5 | - | - | - |
| Ce/Ce* | 1.0858 | 0.06 | 1.0002 | 0.25 | - | - | - |
| Eu/Eu* | 0.7361 | 0.28 | 1.0012 | 0.34 | - | - | - |
| Tm/Tm* | 0.9279 | 0.20 | 0.9981 | 0.51 | - | - | - |

[1] Mean of CI-Chondrites are based on the average of Orgueil, Ivuna and Alais measurements from Pourmand et al. (2012) and this study. These values are used for CI-normalizations and calculations of Ce, Eu and Tm anomalies throughout this contribution.

**Table 2**. Rare earth element concentrations (μg g$^{-1}$), Eu, Ce and Tm anomalies for chondrites and achondrites.

| Meteorite Name | Meteorite Group | Fall | Source | Collection ID | Homogenized (digested) Mass (g) | La | Ce | Pr | Nd | Sm | Eu | Gd | Tb | Dy | Ho | Er | Tm | Yb | Lu | Lu (IDMS)[2] | U (IDMS)[2] | Th (IDMS)[2] | $^{176}$Hf/$^{177}$Hf JMC 475-normalized[2] | Lu/Hf (atomic ratio)[2] | Ce/Ce* | Eu/Eu* | Eu/Eu** | Tm/Tm* | Tm/Tm** | Tm/Tm*** | La$_N$/Lu$_N$ |
|---|---|---|---|---|---|---|---|---|---|---|---|---|---|---|---|---|---|---|---|---|---|---|---|---|---|---|---|---|---|---|
| Alais [1] | CI1 | Y | FM | C3_0067 | 0.20 (0.051) | 0.2670 | 0.6858 | 0.1035 | 0.5299 | 0.1702 | 0.0656 | 0.2318 | 0.0414 | 0.2858 | 0.0612 | 0.1847 | 0.0288 | 0.1882 | 0.0292 | - | - | - | - | - | 1.00 | 1.00 | 0.99 | 1.00 | 0.99 | 0.99 | 0.93 |
| Orgueil A [1] | CI1 | Y | MNHN | 219 | 0.24 (0.076) | 0.2417 | 0.6240 | 0.0947 | 0.4864 | 0.1570 | 0.0601 | 0.2062 | 0.0382 | 0.2636 | 0.0565 | 0.1705 | 0.0263 | 0.1712 | 0.0253 | - | - | - | - | - | 1.00 | 1.01 | 1.02 | 0.99 | 1.00 | 0.98 | 0.98 |
| Orgueil B [1] | CI1 | Y | MNHN | 219 | 0.24 (0.075) | 0.2454 | 0.6242 | 0.0957 | 0.4712 | 0.1513 | 0.0590 | 0.2036 | 0.0375 | 0.2492 | 0.0543 | 0.1611 | 0.0254 | 0.1601 | 0.0241 | - | - | - | - | - | 0.99 | 1.01 | 1.01 | 1.02 | 1.02 | 1.03 | 1.04 |
| Orgueil C [1] | CI1 | Y | MNHN | 219 | 0.24 (0.080) | 0.2521 | 0.6472 | 0.0980 | 0.4966 | 0.1589 | 0.0610 | 0.2142 | 0.0382 | 0.2608 | 0.0554 | 0.1662 | 0.0257 | 0.1681 | 0.0247 | - | - | - | - | - | 1.00 | 1.00 | 1.00 | 0.99 | 1.00 | 0.99 | 1.04 |
| Orgueil D [1] | CI1 | Y | U of Chicago | C3_1146 | 0.90 (0.053) | 0.2506 | 0.6371 | 0.0962 | 0.4843 | 0.1541 | 0.0591 | 0.2078 | 0.0373 | 0.2545 | 0.0547 | 0.1660 | 0.0258 | 0.1719 | 0.0257 | - | - | - | - | - | 1.00 | 1.00 | 0.99 | 0.99 | 0.99 | 0.99 | 1.00 |
| Orgueil E [1] | CI1 | Y | U of Chicago | C3_1146 | 0.90 (0.051) | 0.2454 | 0.6306 | 0.0964 | 0.4909 | 0.1579 | 0.0613 | 0.2138 | 0.0387 | 0.2628 | 0.0568 | 0.1709 | 0.0269 | 0.1737 | 0.0260 | - | - | - | - | - | 1.00 | 1.01 | 1.01 | 1.01 | 1.01 | 1.01 | 0.97 |
| Orgueil H [1] | CI1 | Y | MNHN | 219 | 0.68 (0.096) | 0.3536 | 0.87101 | 0.1256 | 0.6091 | 0.1843 | 0.0668 | 0.2417 | 0.0429 | 0.2914 | 0.0618 | 0.1843 | 0.0288 | 0.1903 | 0.0282 | - | - | - | - | - | 1.01 | 0.96 | 0.95 | 0.99 | 1.00 | 1.00 | 1.28 |
| Orgueil I [1] | CI1 | Y | MNHN | 120 | 0.12 (0.052) | 0.2522 | 0.6498 | 0.0979 | 0.4952 | 0.1584 | 0.0603 | 0.2127 | 0.0378 | 0.2544 | 0.0533 | 0.1574 | 0.0248 | 0.1635 | 0.0243 | - | - | - | - | - | 1.00 | 0.99 | 0.99 | 1.00 | 1.01 | 1.02 | 1.06 |
| Orgueil J [1] | CI1 | Y | MNHN | 120 | 0.12 (0.058) | 0.2496 | 0.6476 | 0.0972 | 0.4928 | 0.1566 | 0.0596 | 0.2107 | 0.0375 | 0.2543 | 0.0538 | 0.1597 | 0.0246 | 0.1642 | 0.0242 | - | - | - | - | - | 1.01 | 0.99 | 0.99 | 0.98 | 0.99 | 0.99 | 1.06 |
| Ivuna A [1] | CI1 | Y | NMNH | 6630 | 0.47 (0.035) | 0.2710 | 0.6894 | 0.1043 | 0.5256 | 0.1682 | 0.0645 | 0.2257 | 0.0407 | 0.2753 | 0.0594 | 0.1787 | 0.0287 | 0.1833 | 0.0300 | 0.02799 | 0.00829 | 0.03185 | 0.282826 | 0.2457 | 1.00 | 1.00 | 1.00 | 1.03 | 1.03 | 1.02 | 0.92 |
| Ivuna B [1] | CI1 | Y | NMNH | 6630 | 0.47 (0.047) | 0.2623 | 0.6724 | 0.1017 | 0.5164 | 0.1653 | 0.0634 | 0.2225 | 0.0396 | 0.2730 | 0.0581 | 0.1762 | 0.0272 | 0.1800 | 0.0263 | - | - | - | - | - | 1.00 | 1.00 | 1.00 | 0.99 | 1.00 | 0.98 | 1.02 |
| Mighei | CM2 | Y | FM | 1456 | 0.46 (0.054) | 0.3650 | 0.9466 | 0.1410 | 0.7133 | 0.2268 | 0.0856 | 0.2910 | 0.0532 | 0.3671 | 0.0810 | 0.2390 | 0.0372 | 0.2414 | 0.0337 | 0.03301 | 0.01064 | 0.03769 | 0.282795 | 0.2421 | 1.01 | 1.01 | 1.02 | 1.00 | 1.02 | 1.03 | 1.11 |
| Murchison A | CM2 | Y | FM | 2640 | 0.96 (0.15) | 0.2175 | 0.6017 | 0.0876 | 0.4424 | 0.1444 | 0.0550 | 0.1796 | 0.0354 | 0.2380 | 0.0496 | 0.1463 | 0.0233 | 0.1572 | 0.0219 | - | - | - | - | - | 1.06 | 1.03 | 1.06 | 0.99 | 1.03 | 1.03 | 1.01 |
| Murchison B | CM2 | Y | FM | 2640 | 0.96 (0.15) | 0.2106 | 0.5746 | 0.0857 | 0.4363 | 0.1425 | 0.0542 | 0.1785 | 0.0357 | 0.2410 | 0.0504 | 0.1487 | 0.0232 | 0.1558 | 0.0221 | - | - | - | - | - | 1.04 | 1.03 | 1.05 | 0.99 | 1.01 | 1.01 | 0.97 |
| Murchison C | CM2 | Y | FM | 2640 | 0.96 (0.15) | 0.2230 | 0.6118 | 0.0891 | 0.4505 | 0.1468 | 0.0557 | 0.1754 | 0.0362 | 0.2412 | 0.0495 | 0.1451 | 0.0232 | 0.1560 | 0.0217 | - | - | - | - | - | 1.05 | 1.05 | 1.09 | 1.00 | 1.04 | 1.04 | 1.05 |
| Kainsaz | CO3.2 | Y | FM | 2755 | 0.41 (0.060) | 0.4729 | 1.1888 | 0.1784 | 0.8904 | 0.2842 | 0.1065 | 0.3700 | 0.0693 | 0.4806 | 0.1018 | 0.3078 | 0.0476 | 0.3105 | 0.0466 | 0.04375 | 0.01373 | 0.04800 | 0.282817 | 0.2397 | 1.00 | 0.99 | 1.00 | 0.99 | 0.99 | 0.99 | 0.98 |
| Lance | CO3.5 | Y | FM | 1351 | 0.32 (0.061) | 0.4255 | 1.1118 | 0.1678 | 0.8533 | 0.2725 | 0.1043 | 0.3626 | 0.0665 | 0.4550 | 0.0968 | 0.2934 | 0.0451 | 0.2976 | 0.0439 | 0.04146 | 0.01618 | 0.04727 | 0.282762 | 0.2385 | 1.01 | 1.00 | 1.00 | 0.99 | 0.99 | 0.98 | 0.99 |
| Vigarano | CV3 | Y | FM | 782 | 0.17 (0.054) | 0.4783 | 1.2263 | 0.1879 | 0.9597 | 0.3029 | 0.1079 | 0.4039 | 0.0769 | 0.5310 | 0.1199 | 0.3705 | 0.0521 | 0.3169 | 0.0512 | 0.04915 | 0.01466 | 0.05349 | 0.282789 | 0.2421 | 0.99 | 0.93 | 0.93 | 0.97 | 0.97 | 0.90 | 0.96 |
| Grosnaja | CV3 | Y | FM | 1732 | 0.23 (0.059) | 0.4882 | 1.1500 | 0.1747 | 0.8747 | 0.2702 | 0.1020 | 0.3517 | 0.0636 | 0.4350 | 0.0928 | 0.2808 | 0.0444 | 0.2906 | 0.0415 | - | - | - | - | - | 0.96 | 1.00 | 1.00 | 1.00 | 1.02 | 1.01 | 1.20 |
| Allende A | CV3 | Y | NMNH 3529 | Split 8, position 5 | 4000 (0.11) | 0.5628 | 1.4066 | 0.2184 | 1.1213 | 0.3562 | 0.1232 | 0.4433 | 0.0801 | 0.5389 | 0.1078 | 0.3135 | 0.0557 | 0.3258 | 0.0450 | - | - | - | - | - | 0.97 | 0.94 | 0.96 | 1.13 | 1.16 | 1.16 | 1.28 |
| Allende B | CV3 | Y | NMNH 3529 | Split 8, position 5 | 4000 (0.082) | 0.5325 | 1.3411 | 0.2060 | 1.0521 | 0.3349 | 0.1133 | 0.4502 | 0.0812 | 0.5498 | 0.1098 | 0.3195 | 0.0546 | 0.3309 | 0.0470 | - | - | - | - | - | 0.98 | 0.88 | 0.88 | 1.09 | 1.11 | 1.11 | 1.16 |
| Allende C | CV3 | Y | NMNH 3529 | Split 8, position 5 | 4000 (0.092) | 0.5249 | 1.3604 | 0.2055 | 1.1507 | 0.3380 | 0.1156 | 0.4239 | 0.0775 | 0.5140 | 0.1036 | 0.3033 | 0.0535 | 0.3258 | 0.0450 | - | - | - | - | - | 1.01 | 0.93 | 0.94 | 1.10 | 1.14 | 1.14 | 1.19 |
| Allende D | CV3 | Y | NMNH 3529 | Split 8, position 5 | 4000 (0.095) | 0.5162 | 1.4007 | 0.2091 | 1.0686 | 0.3409 | 0.1175 | 0.4455 | 0.0892 | 0.5722 | 0.1091 | 0.3140 | 0.0543 | 0.3354 | 0.0463 | 0.04602 | 0.01534 | 0.05819 | 0.282835 | 0.2443 | 1.04 | 0.93 | 0.92 | 1.08 | 1.12 | 1.11 | 1.14 |
| Sahara 97072 | EH3 | N | Private Collection | - | 0.45 (0.059) | 0.2476 | 0.6461 | 0.0970 | 0.4921 | 0.1547 | 0.0604 | 0.2171 | 0.0397 | 0.2750 | 0.0594 | 0.1799 | 0.0265 | 0.1739 | 0.0269 | 0.02527 | 0.00892 | 0.02840 | 0.282748 | 0.2322 | 1.01 | 0.99 | 0.98 | 0.97 | 0.95 | 0.94 | 0.94 |
| Qingzhen | EH3 | Y | R. N. Clayton | - | 0.19 (0.06) | 0.2127 | 0.5704 | 0.0864 | 0.4445 | 0.1408 | 0.0489 | 0.1960 | 0.0371 | 0.2567 | 0.0552 | 0.1651 | 0.0243 | 0.1526 | 0.0242 | 0.02450 | 0.00761 | 0.02584 | 0.282704 | 0.2334 | 1.02 | 0.89 | 0.88 | 0.97 | 0.96 | 0.95 | 0.90 |
| Indarch | EH4 | Y | FM | 3466 | 0.50 (0.057) | 0.2219 | 0.5947 | 0.0806 | 0.3986 | 0.1244 | 0.0492 | 0.1640 | 0.0311 | 0.2096 | 0.0436 | 0.1258 | 0.0187 | 0.1232 | 0.0179 | - | - | - | - | - | 1.08 | 1.04 | 1.04 | 0.97 | 0.98 | 0.99 | 1.27 |
| Indarch | EH4 | Y | FM | 1404 | 0.82 (0.15) | 0.2290 | 0.6055 | 0.0899 | 0.4535 | 0.1452 | 0.0560 | 0.2029 | 0.0364 | 0.2490 | 0.0533 | 0.1596 | 0.0237 | 0.1601 | 0.0245 | 0.02345 | 0.00944 | 0.02737 | 0.282685 | 0.2290 | 1.03 | 0.97 | 0.96 | 0.97 | 0.95 | 0.96 | 0.96 |
| Adhi Kot | EH4 | Y | AMNH | 3993 | 0.61 (0.044) | 0.2020 | 0.5224 | 0.0781 | 0.3977 | 0.1273 | 0.0481 | 0.1758 | 0.0330 | 0.2271 | 0.0485 | 0.1440 | 0.0210 | 0.1364 | 0.0206 | 0.02392 | 0.00786 | 0.02676 | 0.284055 | 0.3329 | 1.01 | 0.97 | 0.96 | 0.97 | 0.95 | 0.95 | 1.00 |
| St. Mark's | EH5 | Y | NMNH | 3027 | 0.47 (0.058) | 0.2214 | 0.5733 | 0.0853 | 0.4289 | 0.1348 | 0.0504 | 0.1859 | 0.0339 | 0.2296 | 0.0465 | 0.1277 | 0.0189 | 0.1205 | 0.0173 | 0.02318 | 0.00839 | 0.02760 | 0.283143 | 0.2697 | 1.01 | 0.96 | 0.95 | 0.98 | 0.95 | 1.03 | 1.31 |
| Saint-Sauveur | EH5 | Y | MNHN | 1456 | 0.21 (0.060) | 0.2184 | 0.5697 | 0.0844 | 0.4275 | 0.1339 | 0.0517 | 0.1829 | 0.0336 | 0.2304 | 0.0492 | 0.1466 | 0.0214 | 0.1403 | 0.0210 | 0.02031 | 0.00762 | 0.02440 | 0.283250 | 0.2724 | 1.02 | 1.00 | 0.99 | 0.96 | 0.95 | 0.95 | 1.06 |
| Yilmia | EL6 | N | FM | 2740 | 0.62 (0.056) | 0.1320 | 0.3784 | 0.0599 | 0.3129 | 0.1024 | 0.0407 | 0.1456 | 0.0274 | 0.1945 | 0.0432 | 0.1356 | 0.0208 | 0.1395 | 0.0229 | 0.02341 | 0.00553 | 0.02305 | 0.282531 | 0.2031 | 1.03 | 1.00 | 0.99 | 0.98 | 0.95 | 0.94 | 0.59 |
| Daniel's Kuil | EL6 | Y | FM | 1500 | 0.26 (0.044) | 0.2952 | 0.8286 | 0.1321 | 0.6784 | 0.2031 | 0.0678 | 0.2706 | 0.0500 | 0.3453 | 0.0744 | 0.2246 | 0.0334 | 0.2214 | 0.0338 | 0.02926 | 0.00814 | 0.02934 | 0.282528 | 0.2161 | 1.02 | 0.87 | 0.86 | 0.97 | 0.95 | 0.95 | 0.89 |
| Blithfield | EL6 | N | FM | 1646 | 0.25 (0.057) | 0.3526 | 1.0399 | 0.1614 | 0.8411 | 0.2650 | 0.0709 | 0.3796 | 0.0705 | 0.4740 | 0.0948 | 0.2548 | 0.0368 | 0.2240 | 0.0318 | 0.03424 | 0.01875 | 0.04831 | 0.284231 | 0.3369 | 1.06 | 0.67 | 0.67 | 0.99 | 1.03 | 1.13 |
| Hvittis | EL6 | Y | FM | 578 | 0.27 (0.055) | 0.2142 | 0.6013 | 0.0960 | 0.5066 | 0.1650 | 0.0538 | 0.2345 | 0.0432 | 0.3023 | 0.0659 | 0.2019 | 0.0301 | 0.2017 | 0.0312 | 0.03038 | 0.00682 | 0.03189 | 0.282651 | 0.2306 | 1.02 | 0.82 | 0.82 | 0.96 | 0.96 | 0.94 | 0.70 |
| Jajh Deh Kot Lalu | EL6 | Y | NMNH | 1260 | 0.43 (0.059) | 0.1372 | 0.3873 | 0.0609 | 0.3179 | 0.1043 | 0.0433 | 0.1459 | 0.0270 | 0.1914 | 0.0425 | 0.1347 | 0.0210 | 0.1503 | 0.0242 | 0.02288 | 0.00490 | 0.02427 | 0.281675 | 0.1468 | 1.03 | 1.06 | 1.05 | 0.96 | 0.95 | 0.95 | 0.58 |
| Eagle | EL6 | Y | FM | 3149 | 0.19 (0.058) | 0.1724 | 0.5098 | 0.0825 | 0.4309 | 0.1414 | 0.0448 | 0.2010 | 0.0383 | 0.2682 | 0.0596 | 0.1852 | 0.0285 | 0.1918 | 0.0308 | 0.03120 | 0.00756 | 0.03105 | 0.281972 | 0.1806 | 1.03 | 0.80 | 0.79 | 0.98 | 0.96 | 0.96 | 0.57 |
| Khairpur | EL6 | Y | FM | 1538 | 0.53 (0.055) | 0.1776 | 0.4570 | 0.0721 | 0.3616 | 0.1142 | 0.0495 | 0.1607 | 0.0302 | 0.2069 | 0.0462 | 0.1421 | 0.0219 | 0.1489 | 0.0241 | 0.02405 | 0.00684 | 0.02577 | 0.282061 | 0.1788 | 0.98 | 1.10 | 1.09 | 0.98 | 0.96 | 0.96 | 0.75 |
| Pillistfer | EL6 | Y | FM | 1647 | 0.34 (0.060) | 0.2488 | 0.6494 | 0.0987 | 0.5055 | 0.1620 | 0.0628 | 0.2186 | 0.0400 | 0.2759 | 0.0590 | 0.1782 | 0.0272 | 0.1796 | 0.0264 | 0.02497 | 0.00582 | 0.02516 | 0.282648 | 0.2260 | 1.01 | 1.01 | 1.01 | 0.98 | 0.99 | 0.98 | 0.97 |
| Pillistfer | EL6 | Y | NHM Vienna | G88 | 0.97 (0.15) | 0.1502 | 0.4321 | 0.0664 | 0.3456 | 0.1137 | 0.0438 | 0.1618 | 0.0292 | 0.1991 | 0.0427 | 0.1293 | 0.0201 | 0.1403 | 0.0224 | - | - | - | - | - | 1.05 | 0.97 | 0.96 | 0.97 | 0.97 | 0.96 | 0.69 |
| Happy Canyon | EL6/7 | N | FM | 2760 | 0.69 (0.061) | 1.3454 | 0.4527 | 0.2106 | 0.8976 | 0.2083 | 0.0537 | 0.2556 | 0.0414 | 0.2719 | 0.0580 | 0.1720 | 0.0253 | 0.1975 | 0.0264 | 0.02312 | 0.18963 | 0.03047 | 0.281758 | 0.1666 | 0.21 | 0.71 | 0.68 | 0.89 | 0.94 | 0.95 | 5.21 |
| Ilafegh 009 | EL7 | N | AMNH | 4757 | 0.28 (0.060) | 0.2060 | 0.4488 | 0.0693 | 0.3570 | 0.1222 | 0.0510 | 0.1644 | 0.0327 | 0.2251 | 0.0486 | 0.1475 | 0.0225 | 0.1614 | 0.0221 | 0.02404 | 0.00548 | 0.02619 | 0.282726 | 0.2446 | 0.92 | 1.09 | 1.10 | 0.96 | 0.95 | 0.96 | 0.88 |
| Bielokrynitschie | H4 | Y | FM | 1394 | 0.55 (0.060) | 0.8830 | 4.0932 | 0.2177 | 0.9029 | 0.2278 | 0.0832 | 0.2819 | 0.0509 | 0.3447 | 0.0739 | 0.2220 | 0.0331 | 0.2223 | 0.0335 | 0.03137 | 0.01217 | 0.04079 | 0.282581 | 0.2218 | 2.29 | 1.00 | 0.97 | 0.96 | 0.96 | 0.96 | 2.70 |
| Ochansk | H4 | Y | FM | 1443 | 0.11 (0.059) | 0.2993 | 0.7891 | 0.1177 | 0.5952 | 0.1894 | 0.0774 | 0.2529 | 0.0479 | 0.3298 | 0.0710 | 0.2135 | 0.0317 | 0.2118 | 0.0321 | 0.03396 | 0.01163 | 0.03946 | 0.282699 | 0.2292 | 1.02 | 1.07 | 1.07 | 0.96 | 0.96 | 0.95 | 0.95 |
| Ochansk | H4 | Y | FM | 1443 | 0.94 (0.16) | 0.2888 | 0.8009 | 0.1213 | 0.6124 | 0.1978 | 0.0758 | 0.2719 | 0.0479 | 0.3163 | 0.0650 | 0.1889 | 0.0287 | 0.1964 | 0.0291 | - | - | - | - | - | 1.04 | 0.99 | 0.98 | 0.96 | 0.97 | 0.99 | 1.02 |
| Kernouve | H6 | Y | MNHN | 602 | 0.27 (0.061) | 0.3202 | 0.8380 | 0.1244 | 0.6295 | 0.2003 | 0.0827 | 0.2609 | 0.0511 | 0.3509 | 0.0758 | 0.2300 | 0.0347 | 0.2312 | 0.0353 | 0.02930 | 0.01203 | 0.03568 | 0.282695 | 0.2319 | 1.02 | 1.09 | 1.10 | 0.96 | 0.95 | 0.96 | 0.93 |
| Kernouve | H6 | Y | MNHN | 602 | 0.78 (0.12) | 0.6809 | 1.3995 | 0.2061 | 0.9611 | 0.2771 | 0.0863 | 0.3686 | 0.0649 | 0.4416 | 0.0941 | 0.2794 | 0.0412 | 0.2776 | 0.0412 | - | - | - | - | - | 0.91 | 0.82 | 0.80 | 0.96 | 0.95 | 0.96 | 1.69 |
| Bald Mountain | L4 | Y | FM | 2392 | 0.45 (0.069) | 0.3759 | 1.0048 | 0.1501 | 0.7560 | 0.2390 | 0.0829 | 0.3138 | 0.0591 | 0.4031 | 0.0862 | 0.2573 | 0.0377 | 0.2492 | 0.0366 | 0.03714 | 0.01045 | 0.03836 | 0.283005 | 0.2573 | 1.03 | 0.91 | 0.92 | 0.96 | 0.96 | 0.95 | 1.05 |
| Barratta | L4 | N | FM | 1463 | 0.16 (0.036) | 0.3051 | 0.8009 | 0.1200 | 0.6073 | 0.1924 | 0.0731 | 0.2587 | 0.0485 | 0.3324 | 0.0715 | 0.2154 | 0.0320 | 0.2156 | 0.0322 | 0.03466 | 0.01350 | 0.03871 | 0.282760 | 0.2377 | 1.02 | 0.99 | 0.99 | 0.96 | 0.96 | 0.95 | 0.97 |
| Dalgety Downs | L4 | Y | FM | 2613 | 0.36 (0.062) | 0.2855 | 0.7116 | 0.1074 | 0.5439 | 0.1721 | 0.0805 | 0.2330 | 0.0434 | 0.3017 | 0.0657 | 0.2013 | 0.0302 | 0.2077 | 0.0320 | 0.03179 | 0.02187 | 0.04911 | 0.282371 | 0.2242 | 0.99 | 1.21 | 1.20 | 0.96 | 0.96 | 0.94 | 0.91 |
| Farmington | L5 | Y | FM | 347 | 0.24 (0.062) | 0.5123 | 1.3461 | 0.1983 | 0.9925 | 0.3098 | 0.0986 | 0.3996 | 0.0762 | 0.5187 | 0.1109 | 0.3320 | 0.0484 | 0.3197 | 0.0468 | 0.04417 | 0.02073 | 0.05382 | 0.283741 | 0.3040 | 1.03 | 0.85 | 0.85 | 0.96 | 0.94 | 0.94 | 1.12 |
| Farmington | L5 | Y | FM | 347 | 0.81 (0.16) | 0.2939 | 0.7947 | 0.1186 | 0.5982 | 0.1934 | 0.0850 | 0.2680 | 0.0484 | 0.3347 | 0.0724 | 0.2193 | 0.0329 | 0.2266 | 0.0348 | - | - | - | - | - | 1.03 | 1.13 | 1.12 | 0.95 | 0.95 | 0.95 | 0.87 |
| Harleton | L6 | Y | FM | 2686 | 0.25 (0.060) | 0.2720 | 0.7247 | 0.1091 | 0.5551 | 0.1786 | 0.0822 | 0.2425 | 0.0454 | 0.3163 | 0.0687 | 0.2101 | 0.0316 | 0.2162 | 0.0331 | 0.03196 | 0.00883 | 0.03456 | 0.282421 | 0.2170 | 1.02 | 1.19 | 1.19 | 0.96 | 0.95 | 0.95 | 0.84 |
| Isoulane | L6 | Y | MNHM | 2268 | 0.87 (0.16) | 0.3026 | 0.8096 | 0.1219 | 0.6134 | 0.1955 | 0.0705 | 0.2685 | 0.0476 | 0.3239 | 0.0689 | 0.2048 | 0.0303 | 0.2048 | 0.0305 | - | - | - | - | - | 1.02 | 0.93 | 0.92 | 0.96 | 0.96 | 0.96 | 1.02 |
| Soko-Banja | LL4 | Y | FM | 1374 | 0.23 (0.060) | 0.3027 | 0.7810 | 0.1173 | 0.5914 | 0.1881 | 0.0819 | 0.2509 | 0.0473 | 0.3265 | 0.0705 | 0.2139 | 0.0320 | 0.2171 | 0.0327 | 0.03383 | 0.01242 | 0.03901 | 0.282722 | 0.2315 | 1.01 | 1.14 | 1.14 | 0.96 | 0.95 | 0.95 | 0.95 |
| Hamlet | LL4 | Y | FM | 3296 | 0.24 (0.058) | 0.3303 | 0.8605 | 0.1285 | 0.6574 | 0.2105 | 0.0896 | 0.2863 | 0.0546 | 0.3734 | 0.0808 | 0.2439 | 0.0372 | 0.2478 | 0.0383 | 0.03901 | 0.01298 | 0.04757 | 0.282619 | 0.2248 | 1.01 | 1.10 | 1.10 | 0.98 | 0.97 | 0.97 | 0.88 |
| Kelly | LL4 | Y | FM | 2235 | 0.19 (0.060) | 0.3404 | 0.8997 | 0.1347 | 0.6806 | 0.2149 | 0.0767 | 0.2957 | 0.0535 | 0.3685 | 0.0790 | 0.2368 | 0.0347 | 0.2296 | 0.0342 | 0.03218 | 0.01507 | 0.03850 | 0.283067 | 0.2751 | 1.02 | 0.92 | 0.91 | 0.95 | 0.95 | 0.95 | 1.02 |
| Kelly | LL4 | Y | FM | 2235 | 0.90 (0.15) | 0.3417 | 0.8940 | 0.1345 | 0.6718 | 0.2118 | 0.0836 | 0.2887 | 0.0508 | 0.3418 | 0.0721 | 0.2135 | 0.0317 | 0.2172 | 0.0327 | - | - | - | - | - | 1.01 | 1.02 | 1.01 | 0.96 | 0.96 | 0.96 | 1.07 |
| Tuxuac | LL5 | Y | FM | 2850 | 0.15 (0.062) | 0.3570 | 0.9315 | 0.1382 | 0.7006 | 0.2244 | 0.0939 | 0.2931 | 0.0577 | 0.3973 | 0.0860 | 0.2616 | 0.0390 | 0.2628 | 0.0401 | 0.03891 | 0.01801 | 0.05003 | 0.282619 | 0.2356 | 1.02 | 1.11 | 1.11 | 0.96 | 0.95 | 0.95 | 0.91 |
| Paragould | LL5 | Y | FM | 2135 | 0.33 (0.060) | 0.5979 | 1.5584 | 0.2306 | 1.1386 | 0.3487 | 0.1042 | 0.4484 | 0.0786 | 0.5680 | 0.1212 | 0.3616 | 0.0521 | 0.3404 | 0.0498 | - | - | - | - | - | 1.02 | 0.80 | 0.80 | 0.96 | 0.96 | 0.94 | 1.23 |
| Paragould | LL5 | Y | FM | 2135 | 0.83 (0.15) | 0.3906 | 1.0382 | 0.1541 | 0.7616 | 0.2402 | 0.0908 | 0.3260 | 0.0584 | 0.3992 | 0.0852 | 0.2547 | 0.0378 | 0.2564 | 0.0386 | 0.04516 | 0.02161 | 0.05841 | 0.283217 | 0.2684 | 1.03 | 0.98 | 0.97 | 0.96 | 0.96 | 0.96 | 1.03 |
| Saint-Séverin | LL6 | Y | MNHN | 2397 | 0.25 (0.059) | 0.7757 | 2.0342 | 0.2993 | 1.4865 | 0.4592 | 0.1054 | 0.6015 | 0.1108 | 0.7529 | 0.1594 | 0.4669 | 0.0664 | 0.4197 | 0.0590 | 0.04270 | 0.01116 | 0.04764 | 0.282735 | 0.2358 | 1.03 | 0.61 | 0.61 | 0.96 | 0.96 | 0.95 | 1.35 |
| Norton County | Aubrite | Y | FM | 2843 #6 | 0.89 (0.11) | 0.0368 | 0.0967 | 0.0140 | 0.0694 | 0.0257 | 0.0038 | 0.0485 | 0.0099 | 0.0744 | 0.0172 | 0.0559 | 0.0092 | 0.0692 | 0.0117 | - | - | - | - | - | 1.04 | 0.32 | 0.30 | 0.96 | 0.95 | 0.97 | 0.32 |
| Bishopville | Aubrite | Y | FM | 251 #9 | 0.81 (0.16) | 0.1308 | 0.2659 | 0.0382 | 0.2001 | 0.0675 | 0.0446 | 0.0969 | 0.0179 | 0.1252 | 0.0273 | 0.0835 | 0.0127 | 0.0873 | 0.0146 | - | - | - | - | - | 0.92 | 1.66 | 1.65 | 0.96 | 0.93 | 0.95 | 0.92 |
| CI | Normalization | | | | | 0.2482 | 0.63662 | 0.0964 | 0.488 | 0.1563 | 0.06 | 0.2102 | 0.0379 | 0.2576 | 0.0551 | 0.1654 | 0.02584 | 0.1686 | 0.02539 | | | | | | | | | | | | |

[1] Data from Pourmand et al. (2012).
[2] Data from Dauphas and Pourmand (2011) by isotope dilution mass spectrometry (IDMS).
All anomalies are calculated relative to the mean of CI-chondrites in Table 1.

**Table 3.** Rare earth element concentrations (µg g$^{-1}$) for terrestrial rocks from Pourmand et al. (2014)[†], Pourmand et al. (2012)[††] and this study[†††].

| Sample ID | La | Ce | Pr | Nd | Sm | Eu | Gd | Tb | Dy | Ho | Er | Tm | Yb | Lu | Dy$_N$/Lu$_N$ | Tm/Tm* | Tm/Tm** | Tm/Tm*** |
|---|---|---|---|---|---|---|---|---|---|---|---|---|---|---|---|---|---|---|
| African Dust[†] | | | | | | | | | | | | | | | | | | |
| 28-30/05/03 | 69.54 | 145.7 | 16.49 | 62.60 | 11.79 | 2.531 | 10.62 | 1.458 | 8.374 | 1.605 | 4.509 | 0.6351 | 4.1824 | 0.5956 | 1.39 | 0.94 | 0.95 | 0.94 |
| 29-30/06/03 | 66.81 | 138.8 | 15.73 | 59.44 | 11.18 | 2.357 | 10.03 | 1.390 | 8.012 | 1.541 | 4.342 | 0.6148 | 4.0813 | 0.5783 | 1.37 | 0.94 | 0.95 | 0.94 |
| 04-06/07/04 | 78.27 | 164.4 | 18.92 | 72.24 | 13.86 | 2.950 | 12.66 | 1.768 | 10.218 | 1.959 | 5.505 | 0.7768 | 5.1144 | 0.7265 | 1.39 | 0.94 | 0.95 | 0.94 |
| 04-08/04/05 | 50.54 | 103.2 | 12.06 | 45.67 | 8.770 | 1.821 | 7.738 | 1.105 | 6.303 | 1.204 | 3.366 | 0.4797 | 3.1531 | 0.4473 | 1.39 | 0.95 | 0.96 | 0.96 |
| 27-31/05/05 | 54.93 | 113.2 | 13.41 | 50.95 | 9.866 | 2.062 | 8.766 | 1.255 | 7.155 | 1.369 | 3.816 | 0.5399 | 3.5233 | 0.5011 | 1.41 | 0.95 | 0.96 | 0.95 |
| 12-15/09/05 | 44.41 | 90.2 | 10.98 | 42.52 | 8.349 | 1.798 | 7.475 | 1.072 | 6.150 | 1.190 | 3.342 | 0.4764 | 3.1234 | 0.4489 | 1.35 | 0.95 | 0.95 | 0.95 |
| 19-20/03/06 | 68.63 | 140.3 | 16.24 | 61.21 | 11.69 | 2.421 | 10.40 | 1.477 | 8.379 | 1.600 | 4.453 | 0.6321 | 4.2047 | 0.5858 | 1.41 | 0.94 | 0.96 | 0.96 |
| 20-21/03/06 | 67.40 | 138.7 | 16.03 | 60.66 | 11.59 | 2.384 | 10.32 | 1.468 | 8.343 | 1.589 | 4.417 | 0.6224 | 4.0991 | 0.5750 | 1.43 | 0.94 | 0.95 | 0.95 |
| 21-22/03/06 | 76.47 | 156.7 | 17.96 | 67.15 | 12.77 | 2.609 | 11.21 | 1.598 | 9.016 | 1.714 | 4.745 | 0.6711 | 4.4627 | 0.6186 | 1.44 | 0.94 | 0.96 | 0.96 |
| 08-12/06/06 | 63.02 | 127.8 | 15.21 | 57.66 | 11.11 | 2.376 | 9.91 | 1.429 | 8.146 | 1.573 | 4.385 | 0.6217 | 4.1249 | 0.5816 | 1.38 | 0.94 | 0.95 | 0.96 |
| 05-09/01/07 | 70.52 | 147.8 | 16.50 | 62.30 | 11.61 | 2.446 | 10.38 | 1.416 | 8.068 | 1.545 | 4.332 | 0.6117 | 4.0479 | 0.5745 | 1.38 | 0.94 | 0.95 | 0.95 |
| 06-11/03/07 | 63.29 | 128.7 | 15.14 | 57.85 | 11.09 | 2.332 | 9.75 | 1.384 | 7.840 | 1.496 | 4.167 | 0.5877 | 3.9251 | 0.5488 | 1.41 | 0.94 | 0.95 | 0.95 |
| 22-26/07/07 | 42.56 | 86.2 | 10.19 | 38.60 | 7.428 | 1.580 | 6.594 | 0.945 | 5.438 | 1.049 | 2.956 | 0.4212 | 2.7910 | 0.3977 | 1.35 | 0.94 | 0.95 | 0.95 |
| 06-09/05/08 | 55.18 | 111.1 | 13.12 | 49.88 | 9.684 | 2.059 | 8.614 | 1.229 | 7.009 | 1.346 | 3.779 | 0.5349 | 3.5431 | 0.5027 | 1.37 | 0.94 | 0.95 | 0.95 |
| 29/08-01/09/08 | 46.18 | 91.2 | 10.69 | 44.42 | 9.073 | 1.925 | 8.006 | 1.151 | 6.684 | 1.282 | 3.607 | 0.5085 | 3.3890 | 0.4771 | 1.38 | 0.94 | 0.95 | 0.94 |
| 25-29/06/09 | 42.04 | 86.4 | 10.44 | 42.98 | 8.727 | 1.889 | 7.751 | 1.112 | 6.436 | 1.230 | 3.440 | 0.4856 | 3.2039 | 0.4544 | 1.40 | 0.94 | 0.95 | 0.95 |
| 04-08/08/09 | 52.88 | 105.4 | 12.52 | 47.51 | 9.072 | 1.917 | 8.039 | 1.154 | 6.709 | 1.288 | 3.623 | 0.5116 | 3.3679 | 0.4788 | 1.38 | 0.94 | 0.95 | 0.94 |
| 26-29/09/09 | 49.99 | 99.0 | 11.73 | 46.35 | 9.178 | 1.945 | 8.111 | 1.151 | 6.615 | 1.265 | 3.543 | 0.4985 | 3.2888 | 0.4657 | 1.40 | 0.94 | 0.95 | 0.94 |
| 30/03-02/04/10 | 52.21 | 102.5 | 12.33 | 48.52 | 9.634 | 1.956 | 8.456 | 1.213 | 6.924 | 1.304 | 3.601 | 0.5023 | 3.2897 | 0.4601 | 1.48 | 0.94 | 0.95 | 0.95 |
| 04-07/04/10 | 52.59 | 105.6 | 12.88 | 48.66 | 9.519 | 1.955 | 8.457 | 1.216 | 6.975 | 1.321 | 3.661 | 0.5105 | 3.3166 | 0.4685 | 1.47 | 0.94 | 0.95 | 0.94 |
| 26-29/04/10 | 49.26 | 96.1 | 12.37 | 47.47 | 9.569 | 2.058 | 8.678 | 1.255 | 7.221 | 1.358 | 3.735 | 0.5283 | 3.4451 | 0.4847 | 1.47 | 0.95 | 0.96 | 0.96 |
| 01-04/06/10 | 56.98 | 110.2 | 13.98 | 52.62 | 10.27 | 2.180 | 9.109 | 1.302 | 7.484 | 1.424 | 3.962 | 0.5567 | 3.6439 | 0.5153 | 1.43 | 0.94 | 0.95 | 0.95 |
| 23-25/05/11 | 59.24 | 116.9 | 14.40 | 53.79 | 10.40 | 2.187 | 9.122 | 1.300 | 7.472 | 1.425 | 3.980 | 0.5601 | 3.7084 | 0.5214 | 1.41 | 0.94 | 0.95 | 0.95 |
| 11-14/08/11 | 50.71 | 97.5 | 12.28 | 46.18 | 9.022 | 1.927 | 8.068 | 1.167 | 6.794 | 1.312 | 3.688 | 0.5187 | 3.3659 | 0.4833 | 1.39 | 0.95 | 0.95 | 0.94 |
| 02-04/11/11 | 51.83 | 100.2 | 12.36 | 46.28 | 8.965 | 1.888 | 7.856 | 1.109 | 6.328 | 1.203 | 3.349 | 0.4709 | 3.1045 | 0.4383 | 1.42 | 0.94 | 0.95 | 0.95 |
| Post Archean Australian Shale (PAAS)[††] | | | | | | | | | | | | | | | | | | |
| AO-6 | 37.87 | 75.23 | 8.596 | 31.46 | 5.728 | 0.990 | 4.911 | 0.720 | 4.273 | 0.8386 | 2.446 | 0.3605 | 2.419 | 0.3544 | 1.19 | 0.96 | 0.96 | 0.95 |
| AO-7 | 40.30 | 78.27 | 8.957 | 32.29 | 5.878 | 1.031 | 5.127 | 0.761 | 4.557 | 0.9096 | 2.687 | 0.3998 | 2.703 | 0.3987 | 1.13 | 0.96 | 0.96 | 0.95 |
| AO-9 | 38.65 | 74.56 | 8.661 | 31.50 | 5.759 | 1.002 | 4.944 | 0.720 | 4.283 | 0.8442 | 2.477 | 0.3649 | 2.459 | 0.3600 | 1.17 | 0.95 | 0.96 | 0.94 |
| AO-10 | 45.69 | 87.48 | 10.14 | 37.01 | 6.935 | 1.264 | 6.102 | 0.861 | 5.027 | 0.9819 | 2.865 | 0.4208 | 2.819 | 0.4119 | 1.20 | 0.96 | 0.96 | 0.94 |
| AO-12 | 44.59 | 82.50 | 10.02 | 37.01 | 6.149 | 1.115 | 5.388 | 0.857 | 5.291 | 1.064 | 3.121 | 0.4561 | 3.064 | 0.4508 | 1.16 | 0.95 | 0.96 | 0.94 |
| SC-7 | 43.08 | 87.61 | 9.90 | 36.75 | 6.835 | 1.211 | 5.983 | 0.867 | 5.135 | 1.009 | 2.911 | 0.4205 | 2.774 | 0.3997 | 1.27 | 0.95 | 0.96 | 0.95 |
| SC-8 | 44.52 | 91.38 | 10.29 | 38.28 | 7.141 | 1.192 | 6.218 | 0.895 | 5.299 | 1.047 | 3.044 | 0.4419 | 2.926 | 0.4254 | 1.23 | 0.95 | 0.96 | 0.94 |
| PL-1 | 42.77 | 90.40 | 10.02 | 36.82 | 7.240 | 1.210 | 6.588 | 1.013 | 6.243 | 1.264 | 3.814 | 0.5781 | 3.916 | 0.5692 | 1.08 | 0.97 | 0.98 | 0.95 |
| PW-5 | 63.53 | 126.8 | 14.72 | 54.78 | 10.287 | 1.917 | 9.127 | 1.328 | 7.816 | 1.515 | 4.314 | 0.6165 | 4.030 | 0.5776 | 1.33 | 0.95 | 0.96 | 0.95 |
| G-3[††] | | | | | | | | | | | | | | | | | | |
| A | 88.74 | 165.23 | 16.55 | 53.652 | 7.049 | 1.3161 | 4.4034 | 0.4538 | 2.0723 | 0.3420 | 0.8708 | 0.1158 | 0.7291 | 0.1025 | 1.99 | 0.93 | 0.93 | 0.94 |
| B | 86.10 | 160.44 | 16.10 | 52.329 | 6.912 | 1.2840 | 4.3357 | 0.4469 | 2.0436 | 0.3354 | 0.8523 | 0.1137 | 0.7102 | 0.1002 | 2.01 | 0.94 | 0.93 | 0.95 |
| C | 89.72 | 167.13 | 16.72 | 54.249 | 7.161 | 1.3375 | 4.5035 | 0.4631 | 2.1069 | 0.3457 | 0.8753 | 0.1154 | 0.7274 | 0.1013 | 2.05 | 0.92 | 0.93 | 0.94 |
| D | 89.36 | 166.03 | 16.70 | 54.051 | 7.029 | 1.3105 | 4.4029 | 0.4502 | 2.0521 | 0.3379 | 0.8607 | 0.1137 | 0.7206 | 0.1013 | 2.00 | 0.92 | 0.93 | 0.94 |
| E | 85.87 | 160.17 | 16.09 | 52.211 | 6.868 | 1.2748 | 4.2994 | 0.4393 | 1.9989 | 0.3286 | 0.8346 | 0.1100 | 0.6947 | 0.0975 | 2.02 | 0.92 | 0.93 | 0.94 |
| F[†††] | 84.82 | 158.48 | 16.07 | 52.441 | 7.038 | 1.3077 | 5.1326 | 0.5073 | 2.0925 | 0.3429 | 0.8585 | 0.1159 | 0.7536 | 0.1028 | 2.01 | 0.92 | 0.94 | 0.97 |
| G-2[††] | | | | | | | | | | | | | | | | | | |
| A | 88.28 | 164.91 | 16.60 | 54.036 | 7.206 | 1.3535 | 4.5144 | 0.4754 | 2.2068 | 0.3641 | 0.9220 | 0.1197 | 0.7467 | 0.1021 | 2.13 | 0.92 | 0.93 | 0.93 |
| B | 88.99 | 166.26 | 16.73 | 54.476 | 7.283 | 1.3656 | 4.5980 | 0.4838 | 2.2439 | 0.3707 | 0.9401 | 0.1253 | 0.7673 | 0.1066 | 2.07 | 0.94 | 0.95 | 0.95 |
| C | 89.31 | 166.41 | 16.69 | 54.109 | 7.096 | 1.3406 | 4.4555 | 0.4622 | 2.1311 | 0.3530 | 0.8978 | 0.1179 | 0.7335 | 0.1015 | 2.07 | 0.93 | 0.93 | 0.94 |
| BCR-2[††] | | | | | | | | | | | | | | | | | | |
| A | 24.50 | 52.61 | 6.675 | 28.366 | 6.441 | 1.9171 | 6.7165 | 1.0288 | 6.3350 | 1.2720 | 3.6415 | 0.5159 | 3.3750 | 0.4967 | 1.26 | 0.95 | 0.94 | 0.94 |
| B | 25.14 | 54.34 | 6.922 | 29.433 | 6.604 | 2.0106 | 6.8628 | 1.0622 | 6.5533 | 1.2965 | 3.6421 | 0.5149 | 3.3180 | 0.4817 | 1.34 | 0.95 | 0.95 | 0.96 |
| BIR-1[††] | | | | | | | | | | | | | | | | | | |
| A | 0.597 | 1.894 | 0.368 | 2.380 | 1.073 | 0.5198 | 1.8206 | 0.3553 | 2.5635 | 0.5642 | 1.7102 | 0.2486 | 1.6423 | 0.2421 | 1.04 | 0.96 | 0.96 | 0.94 |
| B | 0.584 | 1.926 | 0.376 | 2.432 | 1.101 | 0.5341 | 1.9074 | 0.3661 | 2.6400 | 0.5801 | 1.7559 | 0.2561 | 1.6864 | 0.2497 | 1.04 | 0.96 | 0.96 | 0.94 |
| BHVO-1[††] | 15.078 | 38.200 | 5.390 | 25.072 | 6.089 | 2.1028 | 6.3171 | 0.9360 | 5.3938 | 0.9836 | 2.5624 | 0.3320 | 1.9983 | 0.2704 | 1.97 | 0.94 | 0.94 | 0.94 |
| W-2[††] | 10.309 | 23.226 | 2.981 | 13.073 | 3.240 | 1.1093 | 3.7406 | 0.6077 | 3.8872 | 0.7825 | 2.2388 | 0.3170 | 2.0415 | 0.2951 | 1.30 | 0.95 | 0.95 | 0.95 |
| PCC-1[††] | 0.03476 | 0.05783 | 0.00745 | 0.02896 | 0.00612 | 0.00119 | 0.00700 | 0.00121 | 0.01003 | 0.00278 | 0.01155 | 0.00241 | 0.02242 | 0.00454 | 0.22 | 1.00 | 0.97 | 0.96 |
| CI Normalization | 0.248185 | 0.6366235 | 0.09643 | 0.4880498 | 0.1563 | 0.06004 | 0.210174 | 0.037853 | 0.257579 | 0.0551014 | 0.165393 | 0.025836 | 0.168604 | 0.02539 | | | | |

Table 4. Average Tm anomalies in meteorites and terrestrial rocks from this study and Pourmand et al. (2012). Uncertainties are reported at 95% confidence interval of the mean. Outliers were rejected using Chauvenet's criterion.

| Sample Type | Tm/Tm* (%) | ± | Tm/Tm** (%) | ± | Tm/Tm*** (%) | ± |
|---|---|---|---|---|---|---|
| **Meteorites** | | | | | | |
| CI (n=11) | -0.05 | 0.94 | 0.03 | 0.64 | 0.06 | 1.11 |
| CM (n=4) | -0.49 | 1.08 | 2.33 | 1.58 | 2.70 | 1.71 |
| CO (n=2) | -1.02 | - | -0.70 | - | -2.17 | - |
| CV (without Allende, n=2) | -1.09 | - | -2.15 | - | -4.59 | - |
| **Carbonaceous chondrites (n=19)** | **-0.35** | **0.61** | **0.21** | **1.05** | **-0.11** | **1.47** |
| Allende CV3 (n=4) | 10.00 | 3.2 | 13.44 | 3.82 | 12.83 | 3.7 |
| EH (n=7, 6, 6) | -3.11 | 0.83 | -4.21 | 0.94 | -4.44 | 1.70 |
| EH (n=10, 11, 10) | -2.93 | 0.90 | -4.00 | 1.08 | -4.34 | 0.93 |
| **Enstatite chondrites (n=17, 17, 16)** | **-3.00** | **0.56** | **-4.07** | **0.71** | **-4.38** | **0.73** |
| H (n=5, 5, 4) | -3.99 | 0.42 | -4.05 | 0.83 | -4.73 | 0.76 |
| L (n=7) | -4.27 | 0.25 | -4.37 | 0.51 | -5.02 | 0.46 |
| LL (n=7, 8, 8) | -4.16 | 0.28 | -4.03 | 0.61 | -4.71 | 0.92 |
| **Ordinary chondrites (n=19, 20, 19)** | **-4.16** | **0.15** | **-4.15** | **0.30** | **-4.83** | **0.38** |
| Aubrite (n=2) | -3.79 | - | -6.25 | - | -4.37 | - |
| | | | | | | |
| **Terrestrial rocks** | | | | | | |
| African mineral dust (n=25) | -5.84 | 0.14 | -4.76 | 0.12 | -5.10 | 0.23 |
| Post Archaean Australian Shale (n=8, 8, 9) | -4.56 | 0.19 | -3.98 | 0.22 | -5.39 | 0.22 |
| G-3 (granite, n=5, 6, 5) | -7.45 | 0.30 | -6.89 | 0.59 | -6.07 | 1.60 |
| G-2 (granite, n=3) | -6.87 | 2.7 | -6.40 | 2.22 | -5.97 | 2.75 |
| BHVO-1 (basalt, n=1) | -6.44 | - | -5.80 | - | -6.06 | - |
| BCR-2 (basalt, n=2) | -5.06 | - | -5.19 | - | -5.14 | - |
| BIR-1 (basalt, n=2) | -4.25 | - | -4.26 | - | -5.95 | - |
| W-2 (diabase, n=1) | -4.71 | - | -4.56 | - | -5.42 | - |
| PCC-1 (peridotite, n=1) | 0.005 | - | -2.93 | - | -3.88 | - |
| **Terrestrial rocks (n=47, 49, 49)** | **-5.72** | **0.29** | **-4.95** | **0.29** | **-5.36** | **0.19** |

**Table 5. Concentrations (in ppm) of REEs and Hf of phosphates and bulk rocks in ordinary chondrites**

|  | La | Ce | Pr | Nd | Sm | Eu | Gd | Tb | Dy | Ho | Er | Tm | Yb | Lu | Hf |
|---|---|---|---|---|---|---|---|---|---|---|---|---|---|---|---|
| Merrillite (0.34%) | 65.5 | 171.7 | 23.4 | 124.3 | 39.4 | 1.7 | 46.3 | 8.6 | 59.7 | 13.2 | 34.8 | 4.8 | 26.5 | 3.5 | |
| Chlorapatite (0.11%) | 16.4 | 37.7 | 4.5 | 21.1 | 6.1 | 0.8 | 6.3 | 1.2 | 8.0 | 1.7 | 4.5 | 0.6 | 2.9 | 0.4 | |
| Phosphate (0.45%) | 53.5 | 139.0 | 18.8 | 99.0 | 31.3 | 1.5 | 36.5 | 6.8 | 47.0 | 10.4 | 27.4 | 3.8 | 20.7 | 2.7 | 0.24 |
| Non-phosphate | 0.0661 | 0.2548 | 0.0413 | 0.2116 | 0.0529 | 0.0691 | 0.1403 | 0.0221 | 0.1424 | 0.0303 | 0.1134 | 0.0204 | 0.1223 | 0.0203 | 0.1663 |
| Bulk ordinary chondrite | 0.307 | 0.879 | 0.126 | 0.656 | 0.193 | 0.076 | 0.304 | 0.053 | 0.353 | 0.077 | 0.236 | 0.037 | 0.215 | 0.032 | 0.167 |

The volume fractions of merrillite and chlorapatite in ordinary chondrites are from Jones et al. (2014). The REE concentrations in phosphates are from Crozaz et al. (1989).
The Hf concentration in phosphates is from Martin et al. (2013). Bulk concentrations are from Wasson and Kallemeyn (1988). Non-phosphate concentrations were calculated by mass-balance.

**Table 6.** Influence of type-II CAI on CI abundances

| | CI | Type II CAI | [CAI-II]/[CI] | CI* | CI*/CI (%) |
|---|---|---|---|---|---|
| | | | | Tm/Tm**=-4.5 % | |
| H | 1.97 % | - | | 1.97 | |
| He | 0.00917 ppm | - | | 0.00918 | |
| Li | 1.45 ppm | - | | 1.45 | |
| Be | 0.0219 ppm | ? | | 0.0219? | |
| B | 0.775 ppm | - | | 0.776 | |
| C | 3.48 % | - | | 3.49 | |
| N | 0.295 % | - | | 0.295 | |
| O | 45.9 % | 35.3 % | 0.8 | 45.9 % | 0.03 |
| F | 58.2 ppm | - | | 58.3 | |
| Ne | 0.00018 ppm | - | | 0.00018 | |
| Na | 4962 ppm | 7312 ppm | 1.5 | 4958 ppm | -0.07 |
| Mg | 9.54 % | 5.72 % | 0.6 | 9.55 % | 0.06 |
| Al | 0.84 % | 16.9 % | 20.1 | 0.82 % | -2.85 |
| Si | 10.7 % | 10.2 % | 1.0 | 10.7 % | 0.01 |
| P | 985 ppm | - | | 986 | |
| S | 5.35 % | - | | 5.36 | |
| Cl | 698 ppm | - | | 699 | |
| Ar | 0.00133 ppm | - | | 0.00133 | |
| K | 546 ppm | - | | 547 | |
| Ca | 0.911 % | 8.4 % | 9.2 | 0.900 % | -1.22 |
| Sc | 5.81 ppm | 51.7 ppm | 8.9 | 5.74 ppm | -1.18 |
| Ti | 447 ppm | 4065 ppm | 9.1 | 442 ppm | -1.21 |
| V | 54.6 ppm | ? | | 54.6? | |
| Cr | 2623 ppm | 1154 ppm | 0.4 | 2625 ppm | 0.08 |
| Mn | 1916 ppm | - | | 1919 | |
| Fe | 18.66 % | 5.28 % | 0.3 | 18.68 % | 0.11 |
| Co | 513 ppm | 26 ppm | 0.1 | 514 ppm | 0.14 |
| Ni | 1.091 % | - | | 1.093 | |
| Cu | 133 ppm | - | | 133 | |
| Zn | 309 ppm | 1020 ppm | 3.3 | 308 ppm | -0.34 |
| Ga | 9.62 ppm | - | | 9.63 | |
| Ge | 32.6 ppm | - | | 32.6 | |
| As | 1.74 ppm | - | | 1.74 | |
| Se | 20.3 ppm | - | | 20.3 | |
| Br | 3.26 ppm | - | | 3.26 | |
| Kr | 5.22E-05 ppm | - | | 5.23E-05 | |
| Rb | 2.35 ppm | - | | 2.35 | |
| Sr | 7.79 ppm | 114 ppm | 14.7 | 7.63 ppm | -2.04 |
| Y | 1.46 ppm | 2.65 ppm | 1.8 | 1.46 ppm | -0.12 |
| Zr | 3.63 ppm | 8.16 ppm | 2.2 | 3.62 ppm | -0.19 |
| Nb | 0.283 ppm | 2.41 ppm | 8.5 | 0.280 ppm | -1.12 |
| Mo | 0.961 ppm | 5.75 ppm | 6.0 | 0.954 ppm | -0.74 |
| Ru | 0.69 ppm | ? | | 0.69? | |
| Rh | 0.132 ppm | 0.024 ppm | 0.2 | 0.132 ppm | 0.12 |
| Pd | 0.56 ppm | 0.015 ppm | 0.0 | 0.56 ppm | 0.15 |
| Ag | 0.201 ppm | - | | 0.201 | |
| Cd | 0.674 ppm | - | | 0.675 | |
| In | 0.0778 ppm | - | | 0.0779 | |
| Sn | 1.63 ppm | - | | 1.63 | |
| Sb | 0.145 ppm | - | | 0.145 | |
| Te | 2.28 ppm | - | | 2.28 | |
| I | 0.53 ppm | - | | 0.53 | |
| Xe | 1.74E-04 ppm | - | | 1.74E-04 | |
| Cs | 0.188 ppm | - | | 0.188 | |
| Ba | 2.42 ppm | 30.2 ppm | 12.5 | 2.38 ppm | -1.71 |
| La | 0.2414 ppm | 10.74 ppm | 44.5 | 0.23 ppm | -6.49 |
| Ce | 0.6194 ppm | 30.38 ppm | 49.1 | 0.5750 ppm | -7.17 |
| Pr | 0.0939 ppm | 4.55 ppm | 48.4 | 0.0873 ppm | -7.08 |
| Nd | 0.4737 ppm | 20.79 ppm | 43.9 | 0.4434 ppm | -6.40 |
| Sm | 0.1536 ppm | 6.423 ppm | 41.8 | 0.1442 ppm | -6.09 |
| Eu | 0.05883 ppm | 0.6178 ppm | 10.5 | 0.05800 ppm | -1.42 |
| Gd | 0.2069 ppm | 2.949 ppm | 14.3 | 0.2028 ppm | -1.98 |
| Tb | 0.03797 ppm | 0.5084 ppm | 13.4 | 0.03727 ppm | -1.85 |
| Dy | 0.2558 ppm | 2.742 ppm | 10.7 | 0.2521 ppm | -1.45 |
| Ho | 0.05644 ppm | 0.1917 ppm | 3.4 | 0.05624 ppm | -0.36 |
| Er | 0.1655 ppm | 0.2757 ppm | 1.7 | 0.1653 ppm | -0.10 |
| Tm | 0.02609 ppm | 0.8231 ppm | 31.5 | 0.02490 ppm | -4.56 |
| Yb | 0.1687 ppm | 0.9706 ppm | 5.8 | 0.1675 ppm | -0.71 |
| Lu | 0.02503 ppm | 0.0231 ppm | 0.9 | 0.02503 ppm | 0.01 |
| Hf | 0.1065 ppm | 0.136 ppm | 1.3 | 0.1065 ppm | -0.04 |
| Ta | 0.015 ppm | 0.346 ppm | 23.1 | 0.015 ppm | -3.29 |
| W | 0.096 ppm | 0.290 ppm | 3.0 | 0.096 ppm | -0.30 |
| Re | 0.04 ppm | 0.085 ppm | 2.1 | 0.04 ppm | -0.17 |
| Os | 0.495 ppm | ? | | 0.495? | |
| Ir | 0.469 ppm | 0.126 ppm | 0.3 | 0.470 ppm | 0.11 |
| Pt | 0.925 ppm | 0.163 ppm | 0.2 | 0.926 ppm | 0.12 |
| Au | 0.148 ppm | 0.0089 ppm | 0.1 | 0.148 ppm | 0.14 |
| Hg | 0.35 ppm | - | | 0.35 | |
| Tl | 0.14 ppm | - | | 0.14 | |
| Pb | 2.62 ppm | 0.645 ppm | 0.2 | 2.62 ppm | 0.11 |
| Bi | 0.11 ppm | - | | 0.11 | |
| Th | 0.03 ppm | 0.630 ppm | 21.0 | 0.03 ppm | -2.98 |
| U | 0.0081 ppm | 0.089 ppm | 11.0 | 0.0080 ppm | -1.49 |

The CI composition is from Palme et al. (2014). Type II CAI composition is a compilation of values from Grossman and Ganapathy (1976); Mason and Taylor (1982); Tissot et al. (2014). The elements with question marks are quite refractory (Be, V, Ru and Os) are refractory but no data is available in the litterature. CI* was calculated by removing a type II CAI composition to produce a Tm anomaly of -4.5 %.

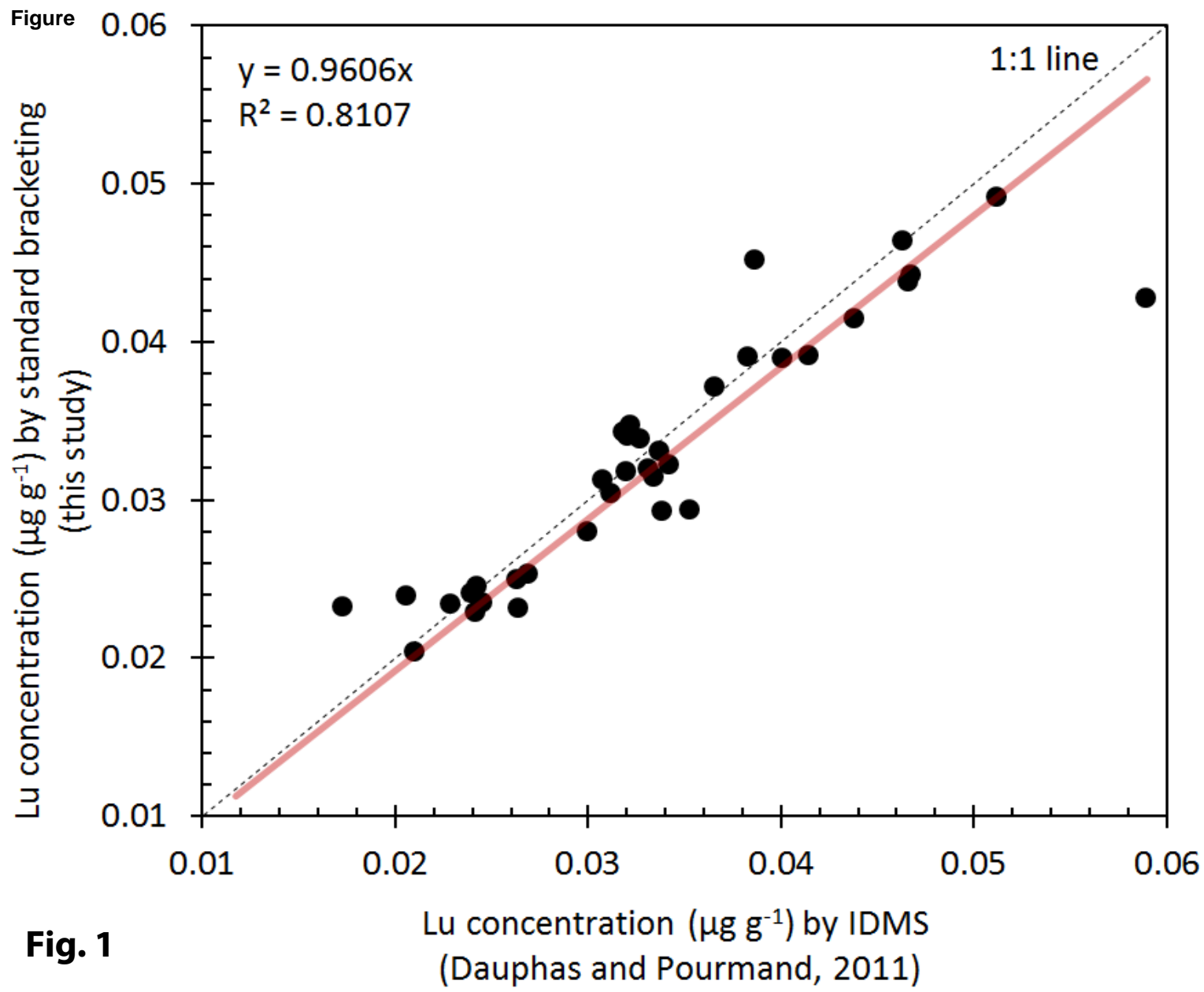

**Fig. 1**

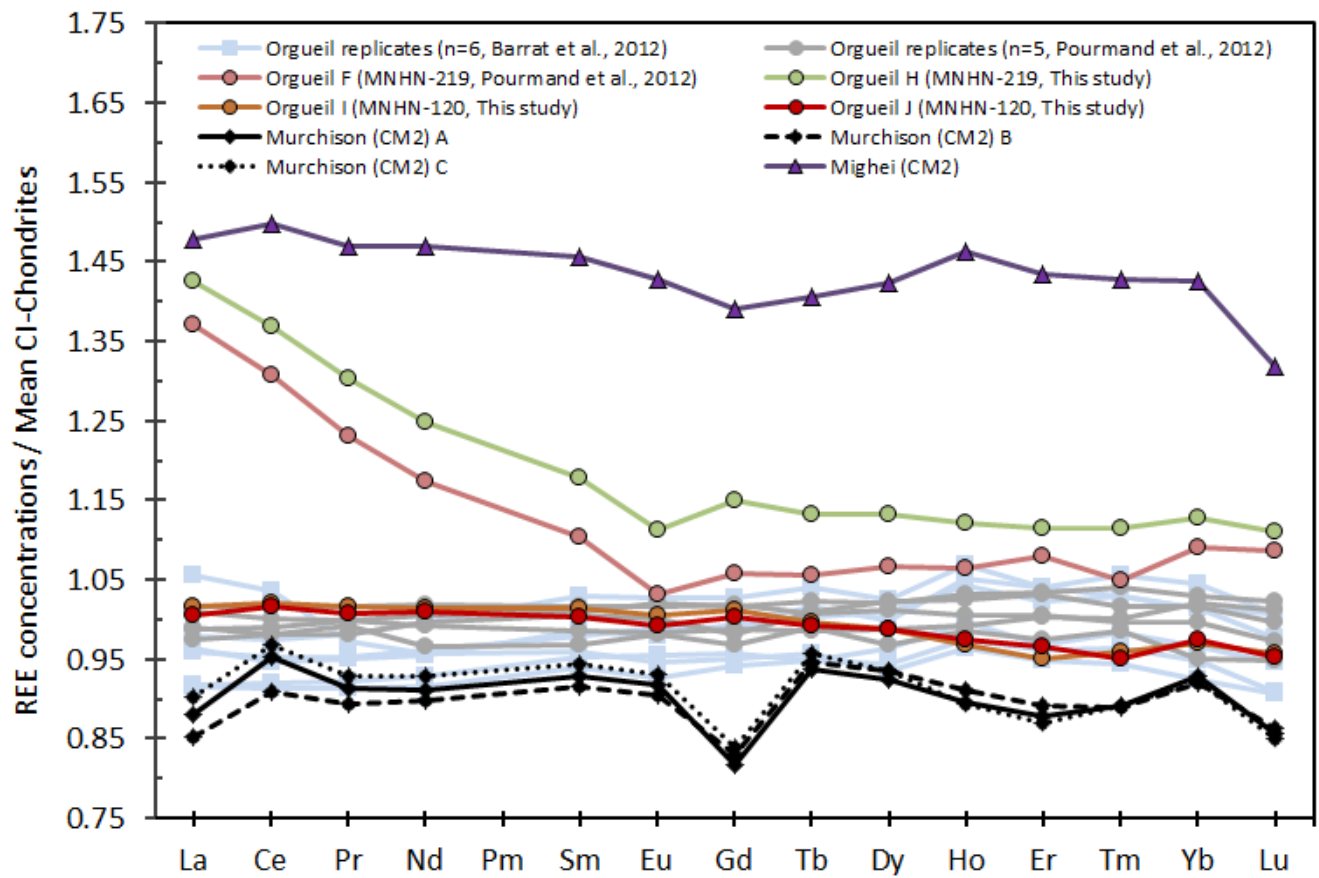

Fig. 2

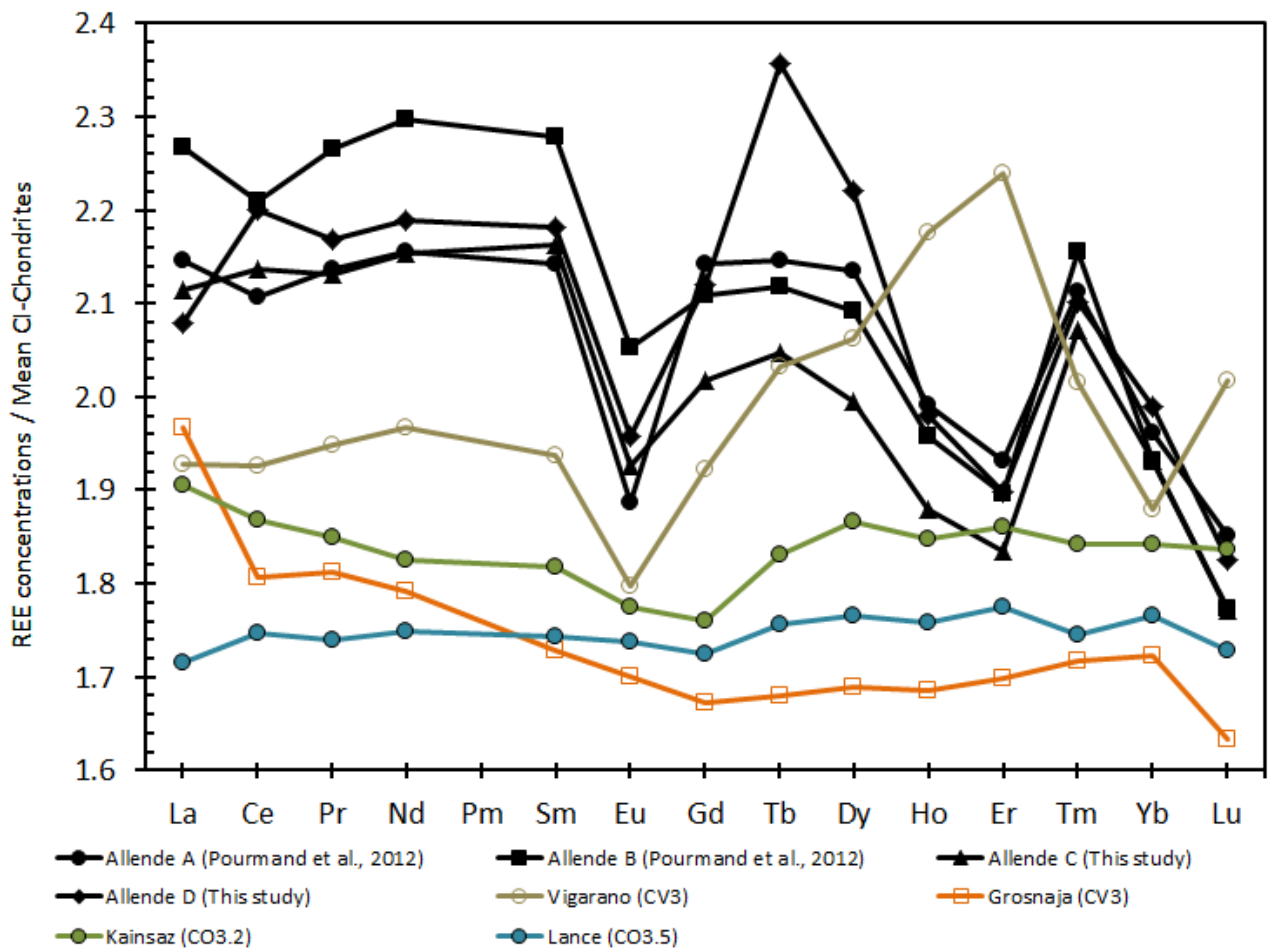

Fig. 3

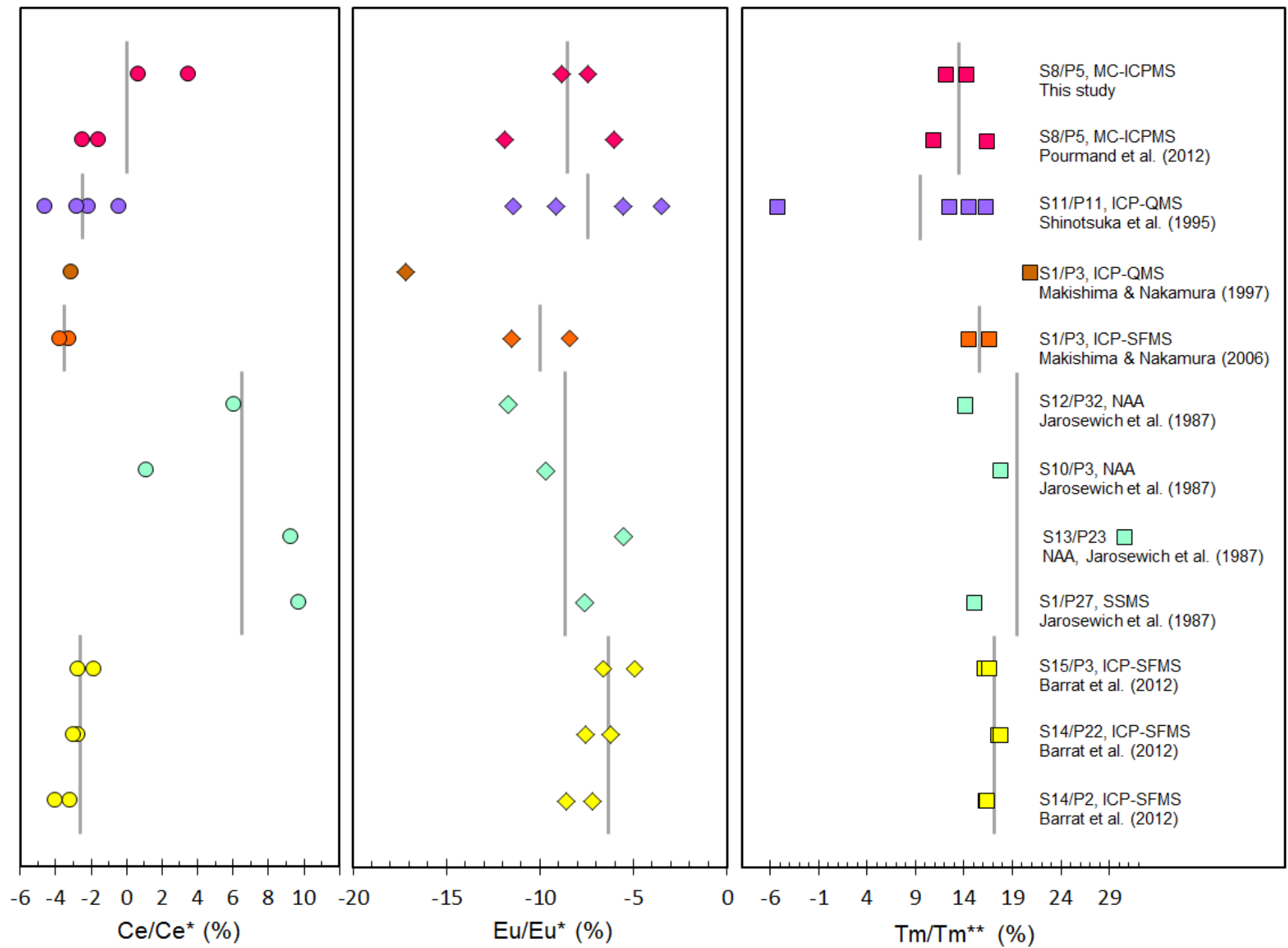

Fig. 4

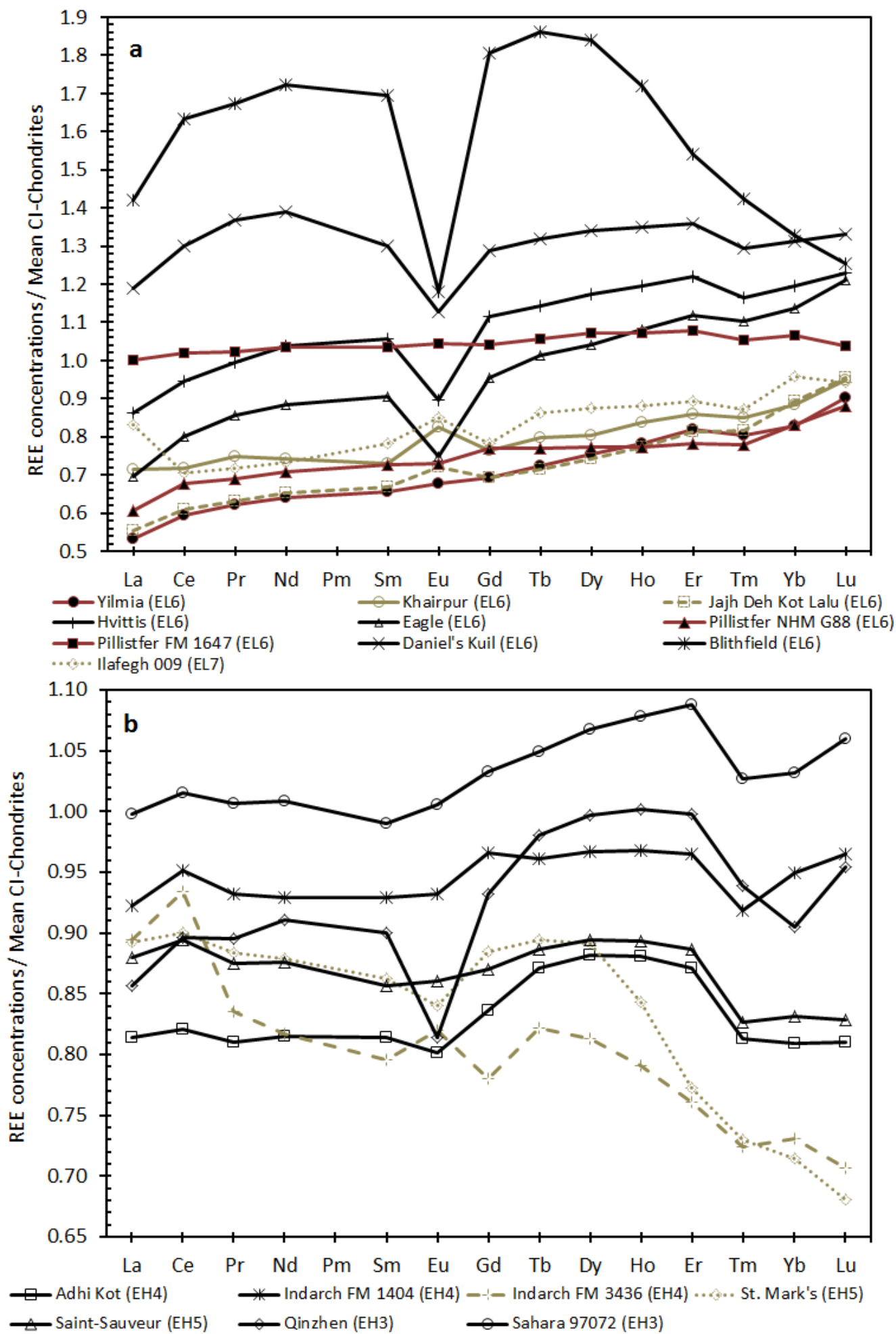

Fig. 5

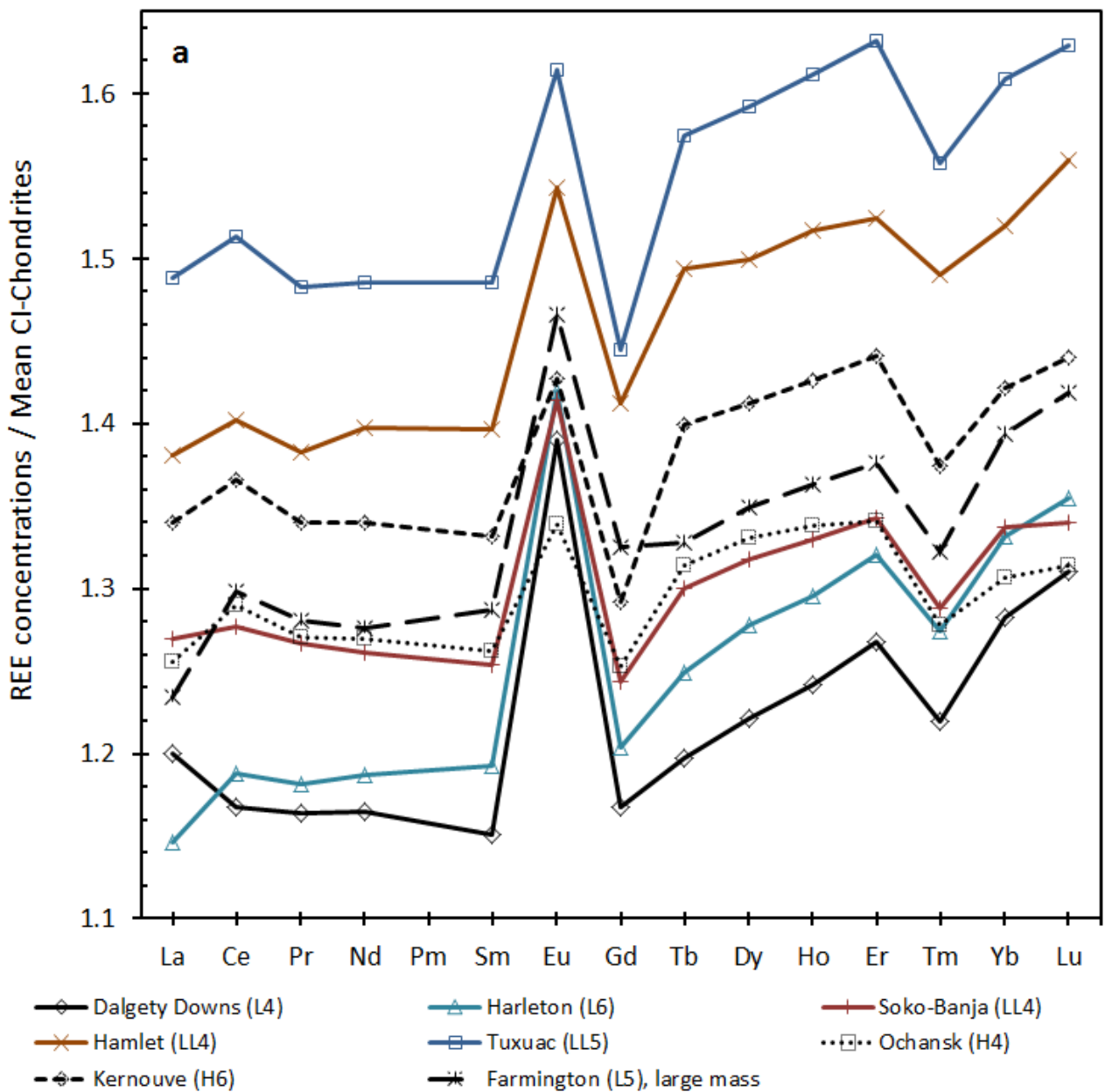
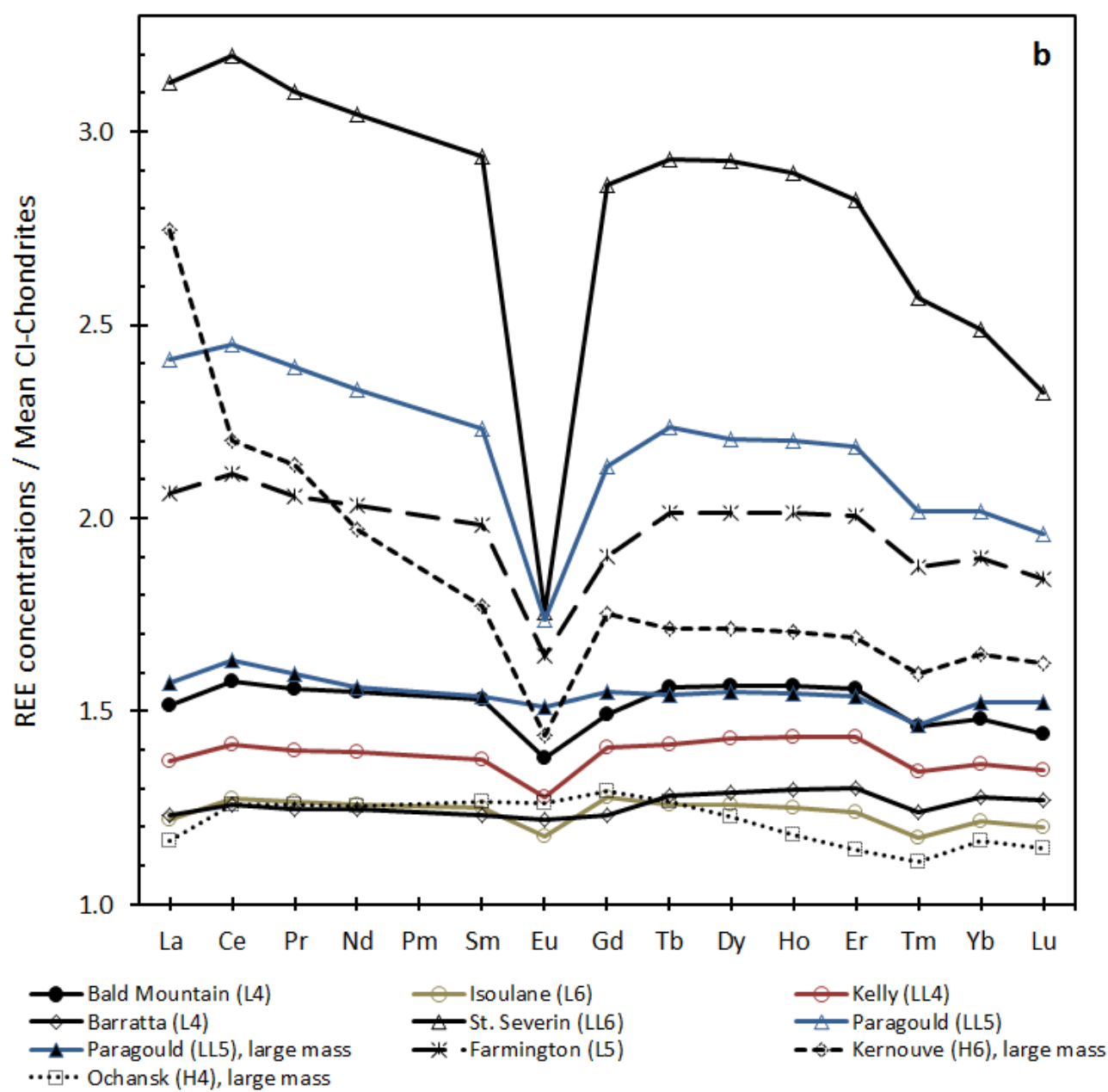

Fig. 6

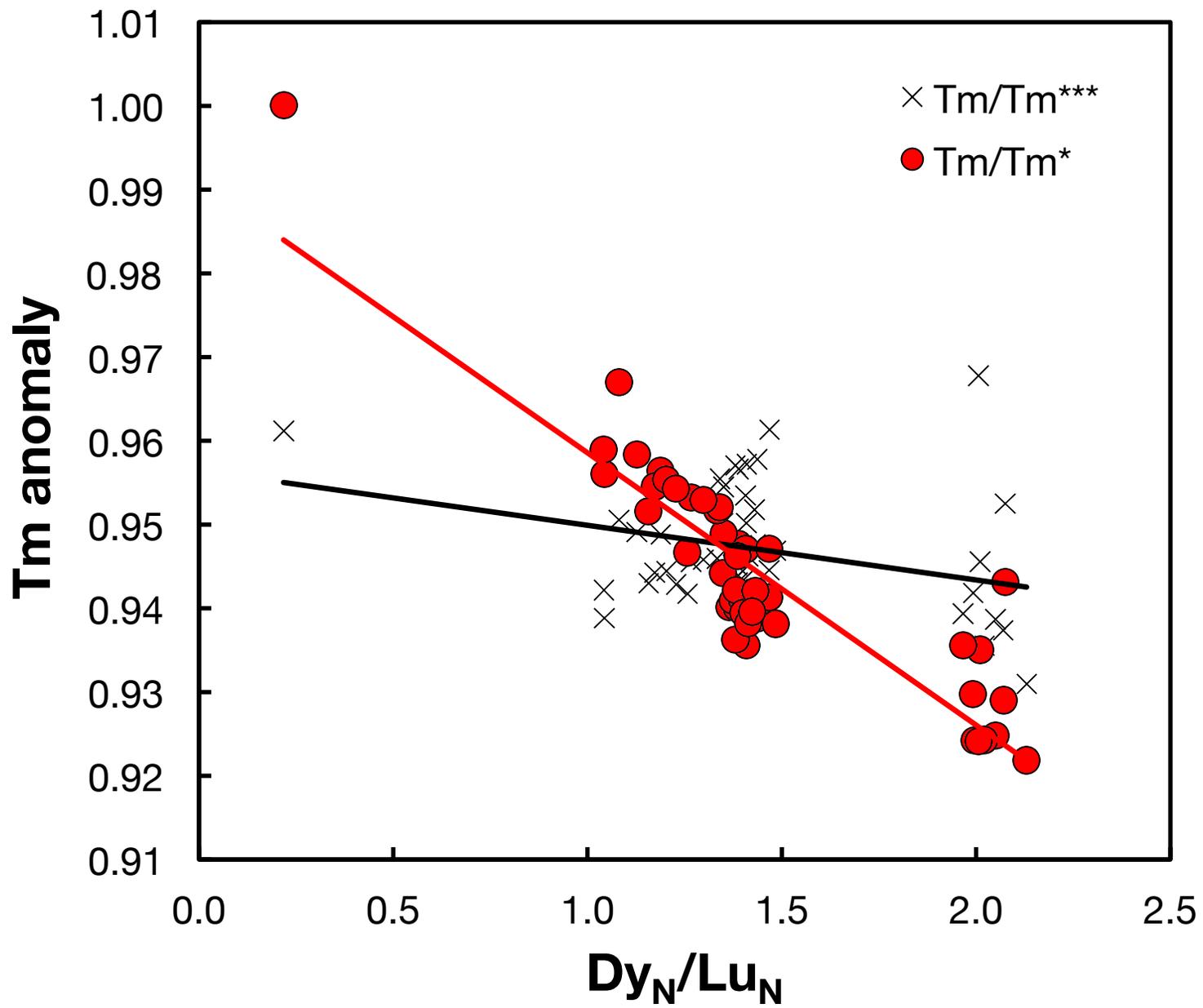

Fig. 7

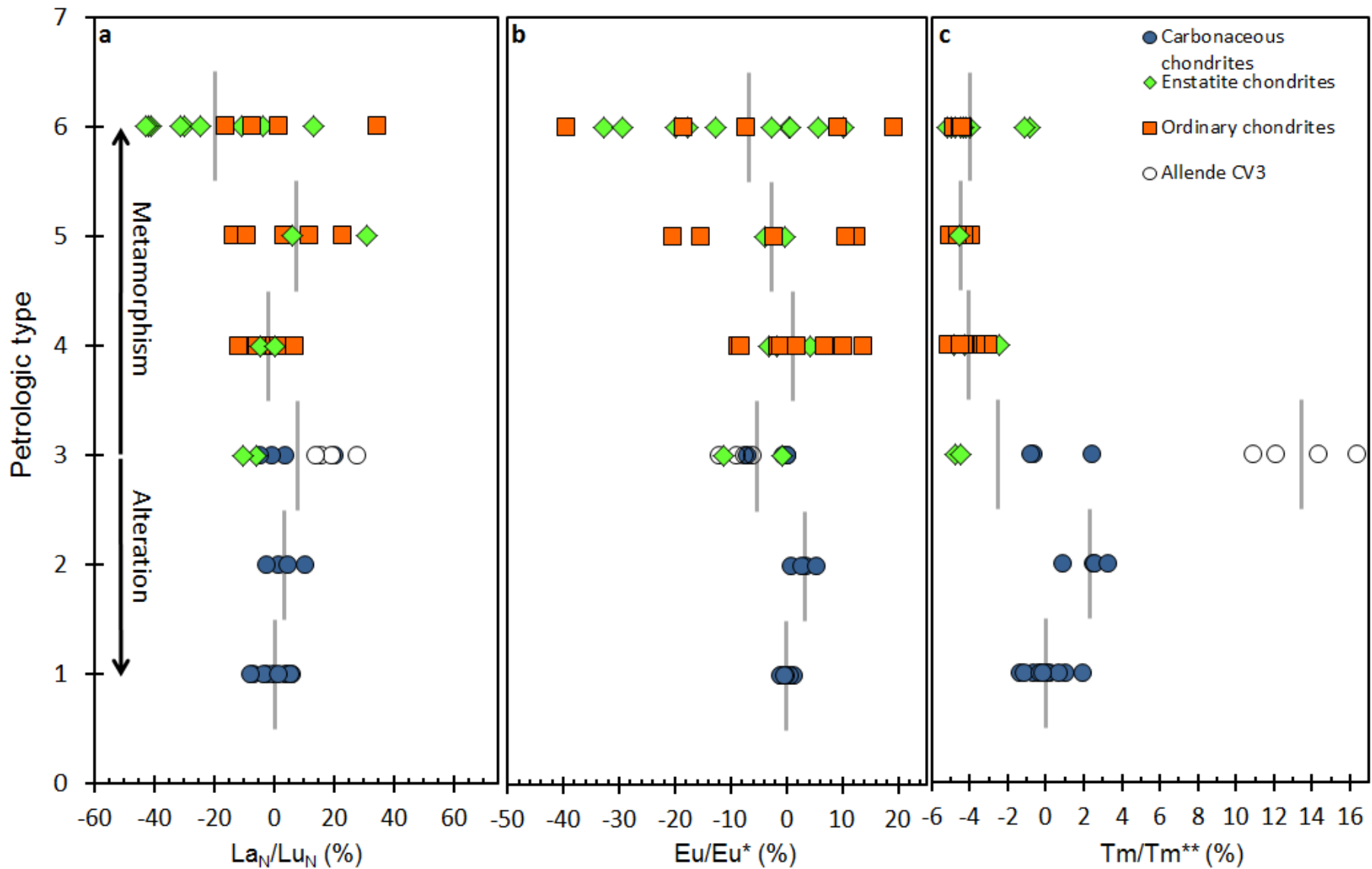

Fig. 8

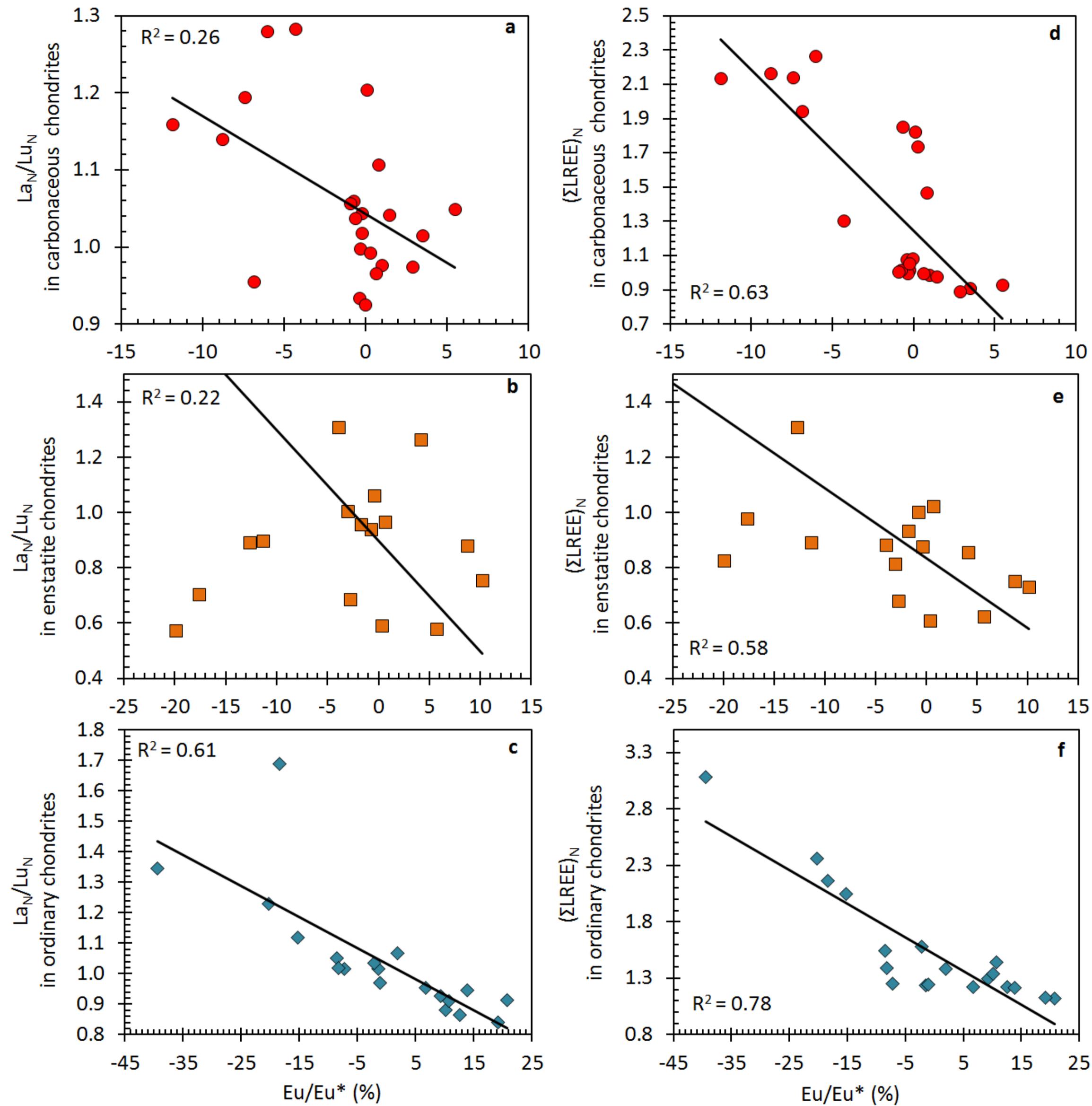

Fig. 9

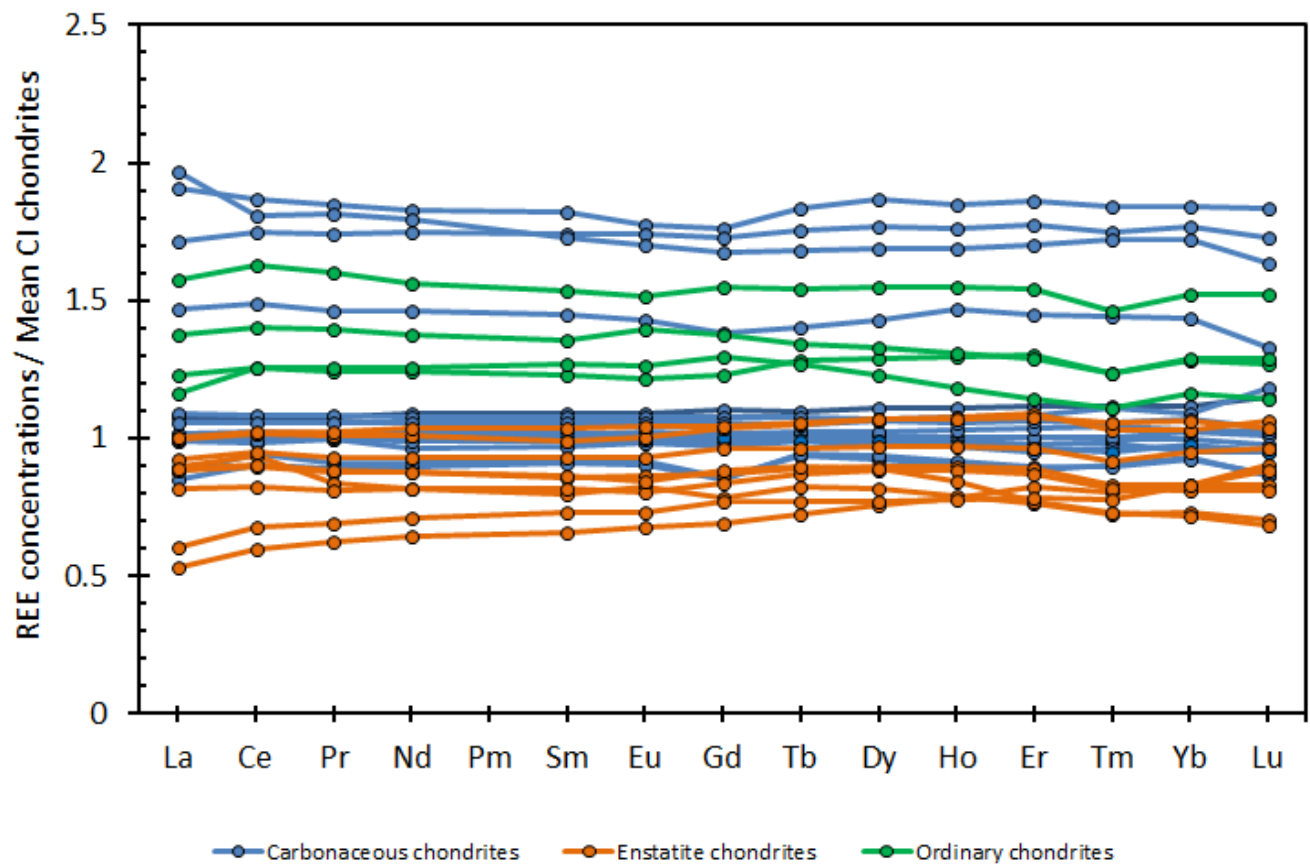

Fig. 10

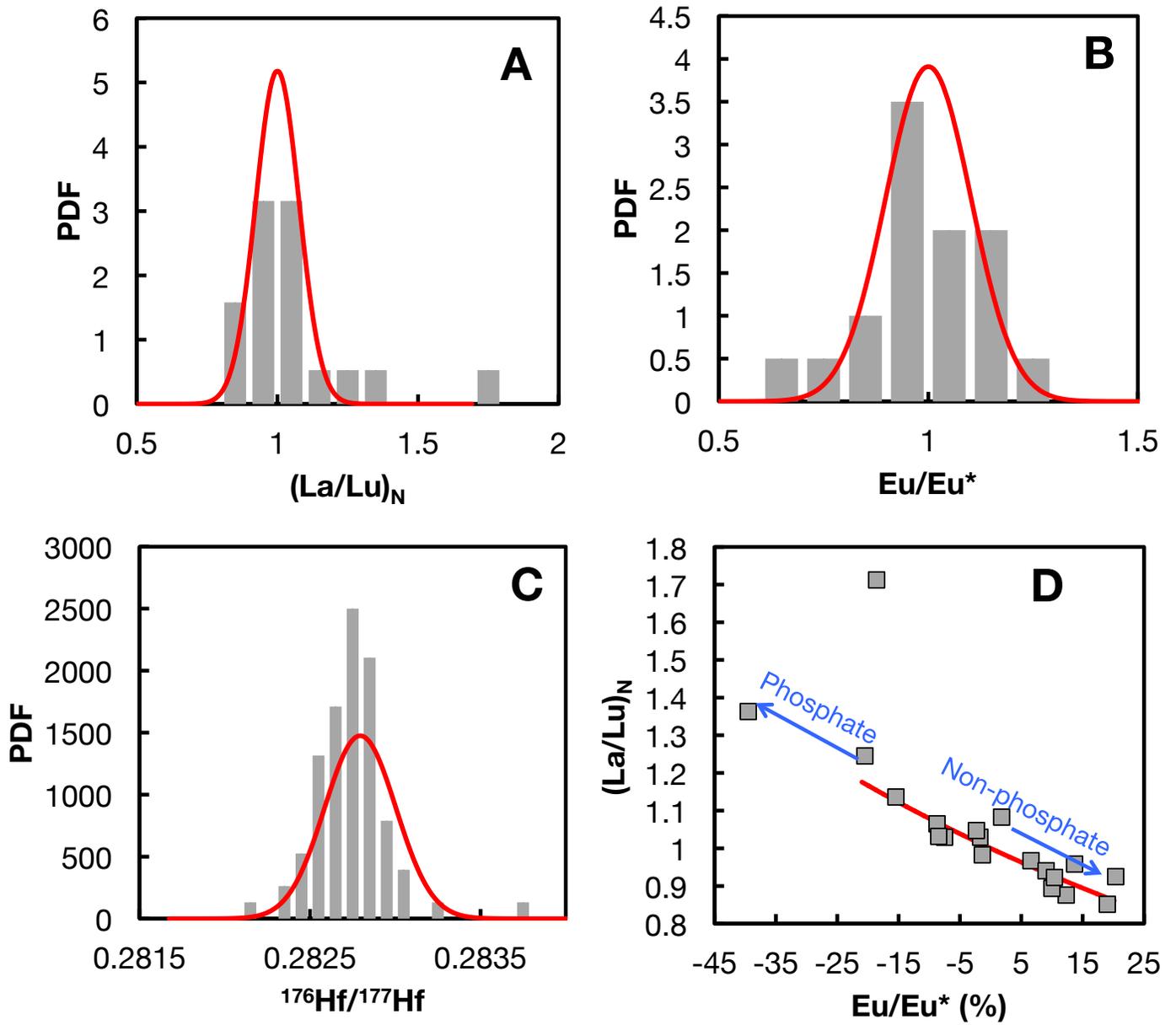

Fig. 11

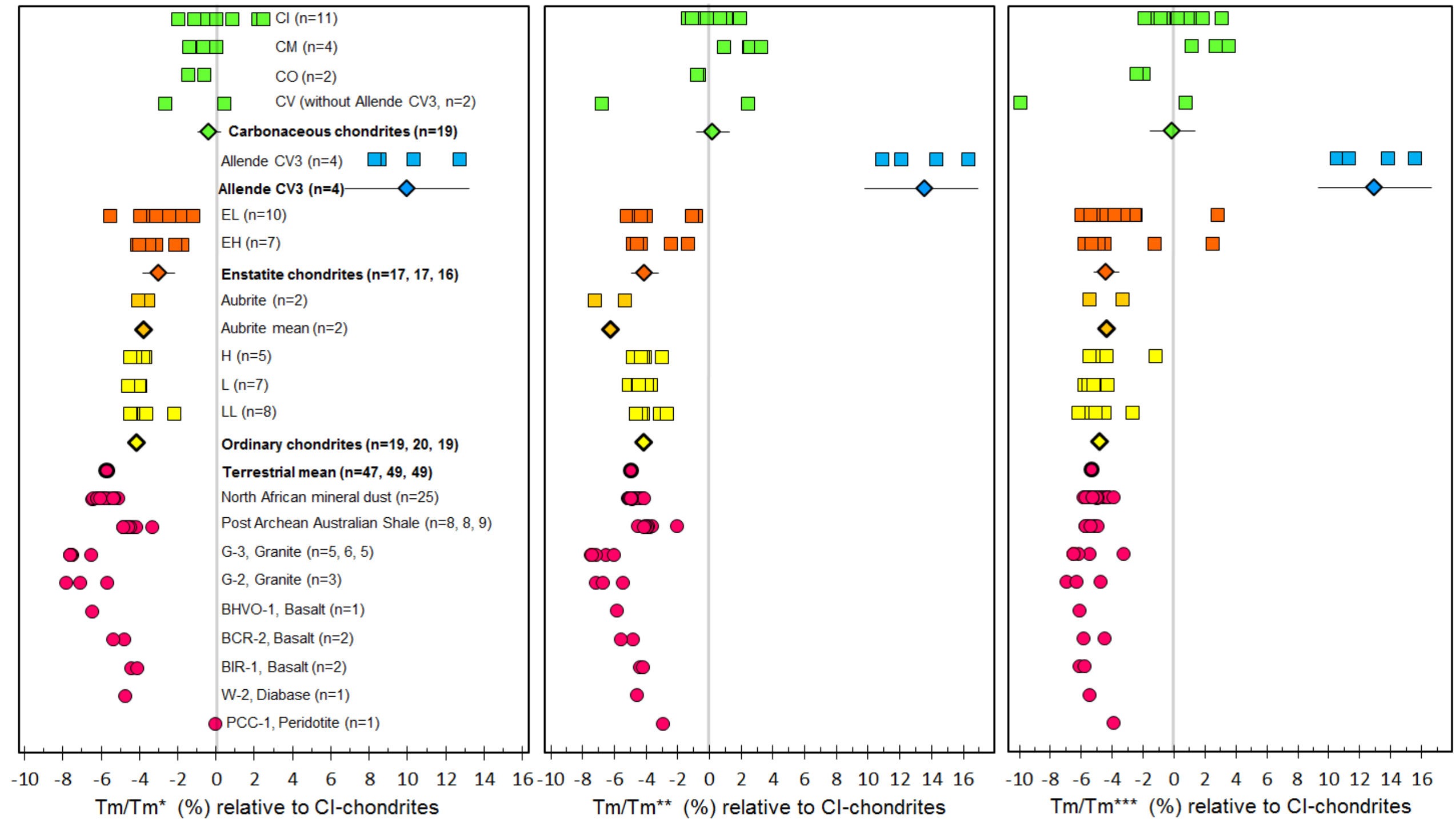

Fig. 12

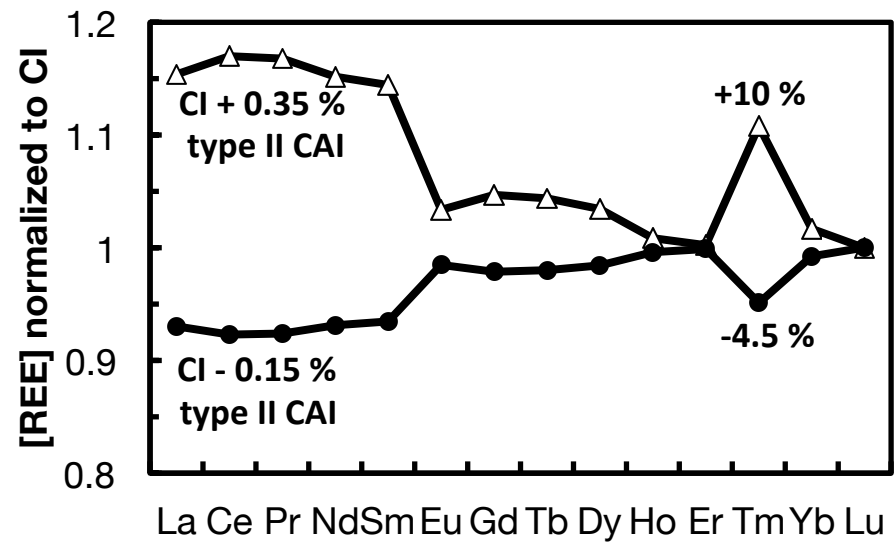

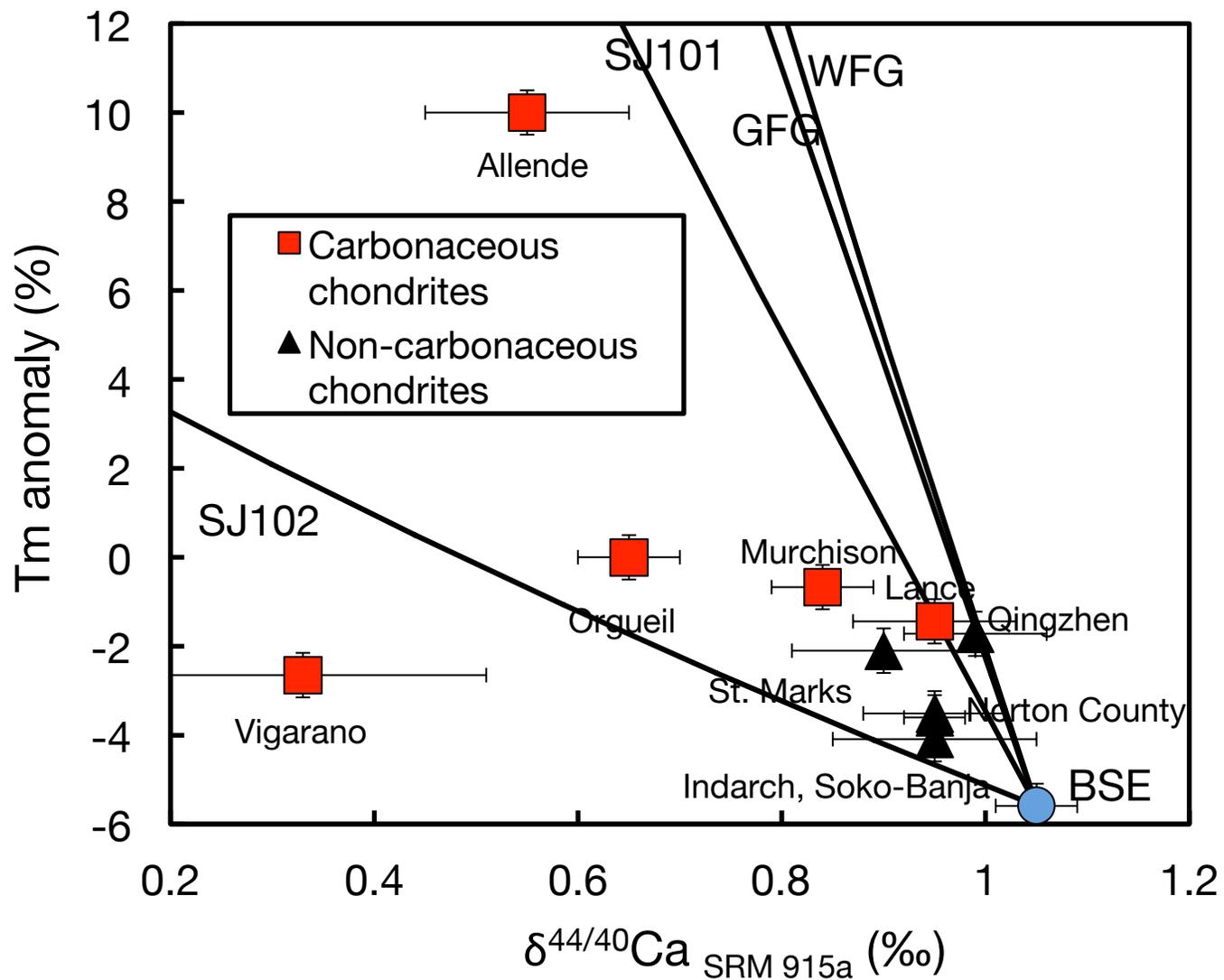